\documentclass[trackchanges]{aastex701}%linenumbers
\usepackage{amsmath}
\usepackage{comment}

\begin{document}

\title{Reexamining Evidence of a Pair-Instability Mass Gap in the Binary Black Hole Population.}%{On the possibility of a pair-instability mass gap in the binary black hole population.}%{A \textit{shallower} look into the high-mass sub-population of binary black holes.}%~$(>40M_{\odot})$ {No evidence for the pair instability mass gap near $40-50M_{\odot}$ in GWTC-4: The astrophysical implications of features in the higher end of the black hole mass-spectrum}

\author[0000-0002-7322-4748]{Anarya\, Ray}
\affiliation{Center for Interdisciplinary Exploration and Research in Astrophysics, Northwestern University, 1800 Sherman Avenue, Evanston, IL 60201, USA}
\affiliation{NSF-Simons AI Institute for the Sky (SkAI), 172 E. Chestnut Street, Chicago, IL 60611, USA}
\email[show]{anarya.ray@northwestern.edu}

\author[0000-0001-9236-5469]{Vicky\, Kalogera}
\affiliation{Center for Interdisciplinary Exploration and Research in Astrophysics, Northwestern University, 1800 Sherman Avenue, Evanston, IL 60201, USA}
\affiliation{NSF-Simons AI Institute for the Sky (SkAI), 172 E. Chestnut Street, Chicago, IL 60611, USA}
\affiliation{Department of Physics and Astronomy, Northwestern University, 2145 Sheridan Road, Evanston, IL 60208, USA}
\email{vicky@northwestern.edu}

\begin{abstract}
% The fourth gravitational wave transient catalog~(GWTC-4) has enabled empirical probes of the theorized pair-instability gap in the higher end of the binary black hole~(BBH) mass-spectrum. In this letter, using flexibly parametrized models, we show that at present there is no evidence of a sharp drop-off in the spectrum of black hole masses near $~40-50M_{\odot}$. We constrain the onset of a potential upper mass gap to be above $(57^{+17}_{-10}M_{\odot})$ and the S factor of $^{12}\mathrm{C}(\alpha, \gamma)^{16}O$ at 300 kev to be $101^{+87}_{-23}\mathrm{keV~barns}$ or lower. We simultaneously characterize the transition in the distribution of BBH mass-ratios, effective aligned and effective precessing spins using our flexible models. From the transitions in our inferred spin and mass-ratio distributions, we find that the high-mass broad-spin sub-population has a significant fraction~($52^{+18}_{-23}\%$) of systems with mass ratios in the range $0.6-1$. This suggests that alternatives to the hypothesis of 2G+1G hierarchical systems dominating BBH formation above $\sim 40-50 M_{\odot}$ are more consistent with the GWTC-4 detection sample. We also demonstrate the effects of strong model assumptions on this inference, which can lead to biased astrophysical interpretation from restricted priors. We note that our results do not exclude that a high-mass gap may be identified as our sample size increases.
The fourth gravitational wave transient catalog~(GWTC-4) has enabled empirical probes of the theorized pair-instability gap in the higher end of the binary black hole~(BBH) mass-spectrum. In this letter, using flexibly parametrized models, we show that at present there is no evidence of a sharp drop-off in the spectrum of black hole masses near $~40-50M_{\odot}$. We simultaneously characterize the transition in the distribution of BBH mass-ratios, effective aligned and effective precessing spins using our flexible models. From the transitions in our inferred spin and mass-ratio distributions, we find that the high-mass broad-spin sub-population has a significant fraction~($52^{+18}_{-23}\%$) of systems with mass ratios in the range $0.6-1$. This suggests that alternatives to the hypothesis of 2G+1G hierarchical systems dominating BBH formation above $\sim 40-50 M_{\odot}$ are more consistent with the GWTC-4 detection sample. By comparing with the predictions of star cluster simulations, we further show that contributions from (2G+2G) systems are not abundant enough to alleviate this discrepancy. We also demonstrate the effects of strong model assumptions on this inference, which can lead to biased astrophysical interpretation from restricted priors. We note that our results do not exclude that a high-mass gap may be identified as our sample size increases. We constrain the lower bound on the location of a possible PISN cutoff still allowed within measurement uncertainties to be $(57^{+17}_{-10}M_{\odot})$ and discuss its implications on the S factor of $^{12}\mathrm{C}(\alpha, \gamma)^{16}O$ at 300 kev.
%By comparing with simulations of dense stellar environments, we show that
%This suggests that alternatives to the hypothesis of 2G+1G hierarchical mergers dominating BBH formation above $\sim 40 M_{\odot}$ are more consistent with the GWTC-4 detection sample.
\end{abstract}


\section{Introduction} 
% \begin{itemize}
%     \item Dearth of m2 in GWTC-4, spin transition in GWTC-4 and previous studies
%     \item astrophysical implications of an evident PISN gap
%     \item alternative explanations
%     \item Model miss-specifcation (Antonini et al do not allow different q distribution above tilde m1), comparison with non-parametric (BGP)
%     \item 
%     \item In this letter 
% \end{itemize}
% Gravitational wave observations of binary black hole mergers can facilitate emperical probes of the theorized Pai
Black holes formed through stellar collapse are theorized to exhibit a gap in their mass spectrum within the range $(40-70M_{\odot}, \sim130M_{\odot})$, due to the complete disruption of progenitors with Helium core masses above $\sim65M_{\odot}$ by (pulsational) pair-instability supernovae~\citep[(P)PISN, ][]{Heger:2001cd, Woosley:2007qp, Belczynski:2016jno, Spera:2017fyx, Leung:2019fgj, Farmer:2019jed, Farmer:2020xne, vanSon:2020zbk, Ziegler:2020klg, Hendriks:2023yrw}. Gravitational wave~(GW) observations of binary black hole~(BBH) mergers by the LIGO-Virgo-KAGRA detector network~\citep[LVK, ][]{KAGRA:2013rdx, LIGOScientific:2014pky,VIRGO:2014yos,KAGRA:2020agh} indicate a substantial population with BH components heavier than $\sim40M_{\odot}$, motivating scenarios that can populate the PISN mass gap~\citep{KAGRA:2021duu, Edelman:2021fik, LIGOScientific:2025pvj}. Dynamically assembled binaries in dense environments such as stellar clusters and disks of active galactic nuclei can contain next-generation BH components that are remnants of previous mergers capable of polluting the PISN deficit~\citep{OLeary:2005vqo, Antonini:2016gqe,Tagawa:2020qll, Mapelli:2021syv, Antonini:2022vib, Torniamenti:2024uxl, Vaccaro:2025ogk, Rodriguez:2019huv}. Identifying such a sub-population in the BBH detection sample can constrain the lower edge of the PISN mass-gap, which remains a central uncertainty in the theoretical predictions of single and binary stellar evolution models.

The location of the PISN cutoff encodes critical information on multiple unknowns in the treatment of massive star evolution such as convective mixing efficiencies, nuclear reaction rates, the role of metallicity and rotation, and various aspects of neutrino physics~\citep{Farmer:2019jed}. Empirical constraints necessitate unambiguous evidence that the observed high mass BBH sub-population is dominated by hierarchical mergers in dense environments. Theoretical descriptions of dynamical interactions in massive stellar clusters predict two key features in this sub-population which are distinct from the rest of the ensemble and therefore can serve as robust observational signatures of an underlying PISN gap.

%Hence, cluster dynamics predicts a unique distribution of BBH effective spin parameters for a purely  hierarchical sub-population that. Therefore, a primary mass-based transtion in the distribution  effective-aligned and effective spin

Firstly, the rarity of next-generation BHs implies a higher rate for hierarchical mergers between a second (2G) and a first (1G) generation companion than all other combinations. Such systems will have significantly greater mass-assymetry than the rest of the ensemble~\citep{Morawski:2018kfs, Gerosa:2018wbw, Rodriguez:2019huv} and populate the PISN deficit in the mass spectrum of the primary (heavier) BH components more efficiently than that of the secondaries, which can still exhibit a sharp cut-off due to PISNe. Secondly, dynamical assembly involves randomization of spin orientation, leading to lower spin-orbit alignment and a preference for highly precessing binaries. Furthermore, as a merger remnant, the 2G component is expected to be highly spinning~\citep[with dimensionless spin-magnitudes of around $\sim0.7$, ][]{Fishbach:2017dwv, Gerosa:2017kvu}. Hence, a primary mass-based transition in the distribution of effective aligned, and effective precessing spin parameters of BBHs into unique shapes predicted by cluster dynamics~\citep{Rodriguez:2019huv,Antonini:2024het} and a simultaneous sharp drop-off in the secondary mass spectrum can serve as a key observational signature of an underlying PISN deficit populated by next-generation components.

Previous studies have attempted to model this signature in the fourth gravitational wave transient catalog~(GWTC-4), which furnishes a large detection sample of the astrophysical BBH population~\citep{LIGOScientific:2025slb}. In a recent study, \cite{Tong:2025wpz} report evidence of a sharp deficit of secondary BHs that onsets at $(m_g=45_{-4}^{+5})$, by modeling this feature explicitly in the mass spectrum of secondary BH components. They further conclude, in agreement with \cite{Antonini:2025ilj}, that the absence of this feature in the mass-distribution of primary components is due to the presence of a sub-population of (2G+1G) hierarchical mergers. They base this conclusion on the observed transition in the distribution of effective aligned spins into a uniform density function, a shape consistent with the predictions for 2G+1G mergers by star-cluster simulations~\citep{Rodriguez:2019huv} and analytical calculations~\citep{Antonini:2024het}, above a primary mass of $\tilde{m}\in(40M_{\odot},50M_{\odot})$. With the complementary measurements of $\tilde{m}$ and $m_{g}$, these studies\footnote{See also, \cite{Afroz:2025ikg}, who explore PISN signatures in GWTC-4 using a phase-space formalism.} have claimed an underlying PISN cut-off in the range $40-50M_{\odot}$ to be evident in GWTC-4 and constrained the carbon-oxygen nuclear reaction rate to be marginally consistent with the independent measurements by \cite{deBoer:2017ldl}.% and the rate of hierarchical (2G+1G)  mergers in stellar clusters. 

However, the inference of population features can be driven by strong prior assumptions~\citep{Mandel:2016prl, Callister:2022qwb, Ray:2025aqr}. For example, \cite{Tong:2025wpz} assume a priori, that the gap in secondary components is wider than $20M_{\odot}$, thereby restricting their inference to a specific set of plausible formation scenarios for high mass BBHs. In this study, using flexibly parametrized models for the BBH mass-ratio distribution, we show that there is no evidence for a sharp cut-off in the secondary mass-spectrum of GWTC-4 BBHs near $(40-50M_{\odot})$ and that sub-populations above the spin-transition mass have a stronger preference for symmetric mass binaries than predicted by theoretical models of hierarchical BBH formation. Our flexible inference is preferred by the data over strongly-prescribed models that either enforce or explicitly model a gap-like feature in the inferred distributions through strong priors. Measurement uncertainties indicate that the current detection sample can still allow for a PISN gap at higher masses, whose lower bound we constrain to be $57^{+17}_{-10}M_{\odot}$ or higher. This implies an S factor of $^{12}\mathrm{C}(\alpha, \gamma)^{16}O$ at 300 kev that is lower than or equal to $101^{+87}_{-23}\mathrm{keV~barns}$, which completely overlaps with the bounds reported by \cite{deBoer:2017ldl} from independent measurements. Our findings motivate alternatives to the hierarchical merger scenario being dominant for BBH formation in the range~$(40-60M_{\odot})$ given GWTC-4 data.% \textcolor{red}{comparison with BGP}

This letter is organized as follows. In Sec.~\ref{sec:pop-inference}, we summarize the details of our population inference. In Sec.~\ref{sec:mass-dist} we show our mass-distribution inference, the lack of evidence in favour of a steep gap. In Sec.~\ref{sec:spin-dist} we perform a spin transition study and show that the post-transition sub-population is likely inconsistent with the 2G+1G hierarchical merger scenario. In Sec.~\ref{sec:model-comparison}, we perform a model comparison analysis and show that a gap can be enforced in the inferred mass-distribution through strong model assumptions and that such an analysis is disfavoured by the data. Finally, we conclude in Sec.~\ref{sec:conclusion}, with a summary of alternative astrophysical explanations for this high mass BBH sub-population in GWTC-4.
\section{Population Inference}
\label{sec:pop-inference}
We model the distribution of BBH primary masses~$(m_1)$, mass ratios~$(q)$, effective inspiral spins~$(\chi_{eff}, \chi_p)$, and redshift~$(z)$. For the primary mass distribution and redshift evolution of the merger rate, we use the same models used by \cite{LIGOScientific:2025pvj}, i.e., a broken power law with two peaks (\texttt{BPL2P}) for the primary and power law in $(1+z)$, and develop flexible parametrizations for the distributions of mass-ratio and effective spin parameters to investigate the presence of a 2G+1G BBH sub-population. We use Bayesian hierarchical inference to constrain our models with the standard inhomogeneous Poisson process likelihood given the parameter estimation~(PE) posterior samples of 153 GWTC-4 events found with a false-alarm rate of one per year or lower, and use ranked simulated signals~(injections)  to correct for Malmquist biases~\citep{Mandel:2018mve, pop-vitale, popgw2, popgw3}. 

We use nested sampling techniques~\citep{Skilling:2006gxv} to constrain the posterior distribution of population hyperparameters and compute the Bayesian evidence in favour of different model assumptions. We monitor the convergence of Monte Carlo sums necessary for evaluating the population likelihood during posterior sampling using the variance-based likelihood penalties proposed by \cite{Talbot:2023pex}. We use the \textsc{gwpopulation} library for implementing our inference~\citep{Talbot2025, 2019PhRvD.100d3030T}. The data of PE and injections are obtained from LVK's public data releases on zenodo~\citep{ligo_scientific_collaboration_2025_16740128, ligo_scientific_collaboration_and_virgo_2021_5546663, ligo_scientific_collaboration_and_virgo_2022_6513631, ligo_scientific_collaboration_and_virgo_2025_16053484}. The code used in this study will be made available along with the inference results upon reasonable request.% at XXXXXX, post manuscript acceptance, and will be available upon request in the meantime.

\section{Mass distribution: Evidence for a smooth fall-off in the secondary BH mass spectrum near $40M_{\odot}$}
\label{sec:mass-dist}
\begin{figure*}[htt]
\begin{center}
\includegraphics[width=\textwidth]{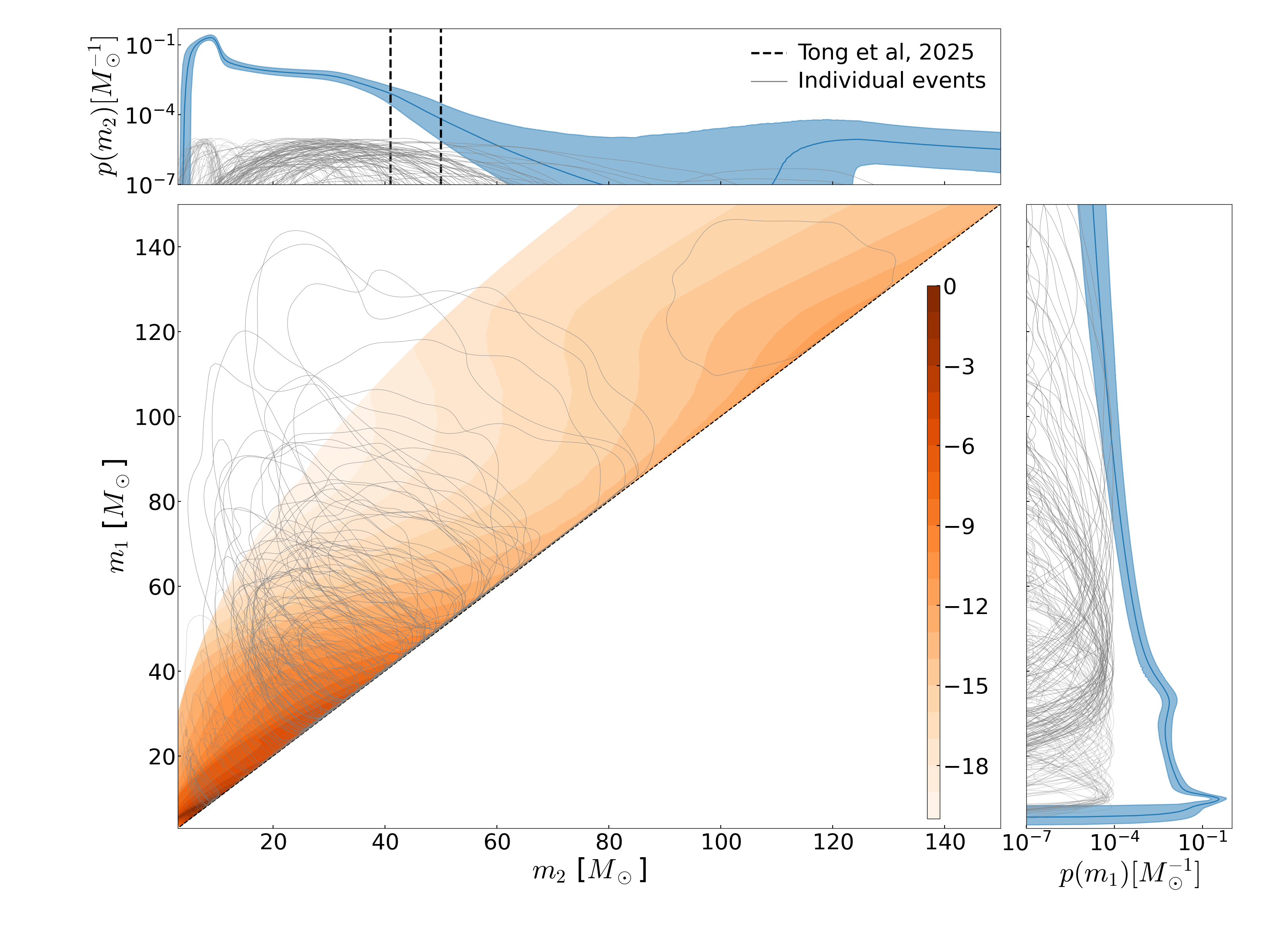}
\caption{\label{fig:pm1m2} Inferred mass distributions show no evidence for a sharp gap. The blue shaded region show demarcates the $90\%$ credible interval obtained from the posterior and the black dashed lines represent the lower edge of a sharp gap inferred by \cite{Tong:2025wpz}.}
\end{center}
\end{figure*}
To investigate the existence of an upper mass gap in the secondary BH components, we first ignore the possibility of spin-transition and assume the distribution functions of BH spin-parameters to be similar to the default model of \cite{LIGOScientific:2025pvj}. This is the same as in the mass gap study of \cite{Tong:2025wpz}\footnote{The default model assumes independent Gaussians for component spin magnitudes and independent mixtures of uniform and Gaussian distributions for cosine tilt angles}. For the mass ratio distribution,  we adopt a more flexible alternative to the parametrizations of \cite{Tong:2025wpz}. Our model is free of sharp features and strong shape priors (especially gap-like priors), reducing the possibility of model-induced biases in the inference. We choose our mass-ratio distribution as:
\begin{equation}
    p(q|m_1) = (1-\lambda_q) \mathcal{TN}_q(q, m_1, \mu_{m_2} \sigma_{m_2}, m_{2,b2}) + \lambda_q \times \mathcal{BPL}_q(q, m_1, m_{2,b1}, \beta_1, \beta_2, m_{2,min}),\label{eq:q-dist}
    %\begin{cases}\frac{q^{\beta_1}(1+\beta_1)}{1-(\frac{m_{2,min}}{m_1})^{\beta_1+1}}, & m_{2,min}<q  m_1 < m_1 < m_{2,b}\\
     %   \frac{q^{\beta_1}(1+\beta_1)}{(\frac{m_{2,b}}{m_1})-(\frac{m_{2,min}}{m_1})^{\beta_1+1}}, & m_{2,min}<q  m_1 <  m_{2,b} < m_1\\
      %  \frac{q^{\beta_2}(1+\beta_2)}{1-(\frac{m_{2,b}}{m_1})^{\beta_1+1}}, & m_{2,b}<q  m_1 < m_1 \\
       % 0, & o.w.\end{cases}
\end{equation}
where $m_2=qm_1$ is the secondary mass, $\mathcal{TN}_q$  a truncated Gaussian, and $\mathcal{BPL}_q$ a \textit{continuous} broken power-law distribution. For the exact functional forms of these distributions, see appendix~\ref{secc:appendix-dist}. See also the findings of \cite{Banagiri:2025dmy}, who identify three distinct sub-populations of BBHs across the entire mass-spectrum that exhibit mass-ratio transitions consistent with a smooth $m_2$ distribution.
%Note that in constrast to~\cite{Tong:2025wpz} who explicitly model an upper-mass gap in $p(m_2)$ and make prior assumptions about its width to be $>20M_{\odot}$, our distribution functions 

The location of the break in terms of secondary mass~($m_{2,b1}$), the post-break powerlaw index ($\beta_2$), and the lower truncation of the Gaussian (also in terms of secondary mass, $m_{2,b2}$), are the key hyperparameters that will be informed by features in the higher end of the secondary component's mass-spectrum. In the limits $\beta_2\rightarrow -\infty$ and $m_{2,b2}>m_{2,b1}$, our model admits a sharp gap spanning the range $m_2\in (m_{2,b1},m_{2,b2})$. In other words, strong priors on these hyperparameters can be chosen to enforce a mass gap. We demonstrate in Sec.~\ref{sec:model-comparison} how such priors are not preferred by the data over the flexible ones but can still be preferred over models that impose a single power-law on the mass-ratio distribution independent of primary mass. For a full list of priors on the population hyperparameters, see appendix~\ref{sec:appendix-priors}.
\begin{figure*}[htt]
\begin{center}
\includegraphics[width=0.31\textwidth]{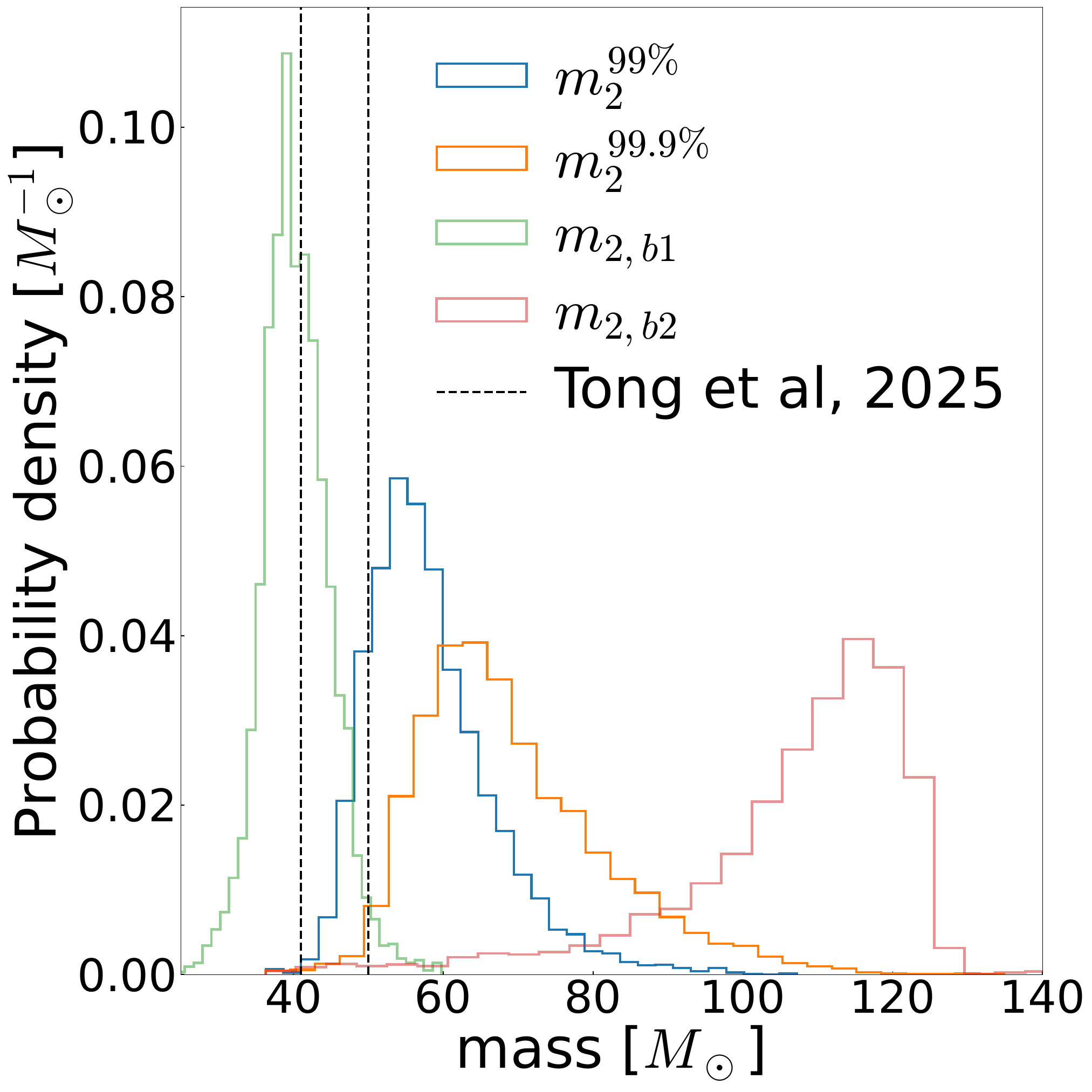}
\includegraphics[width=0.31\textwidth]{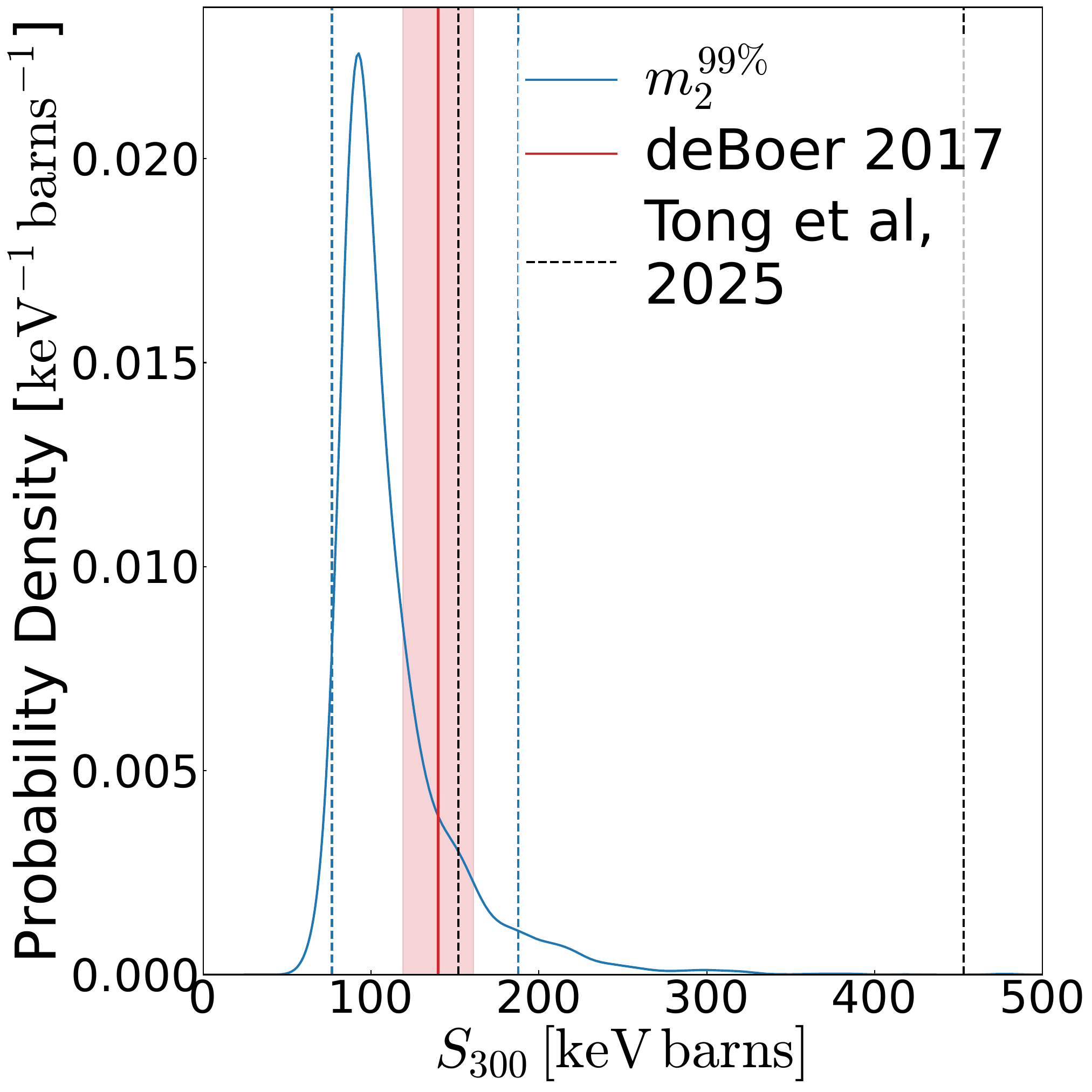}
\includegraphics[width=0.31\textwidth]{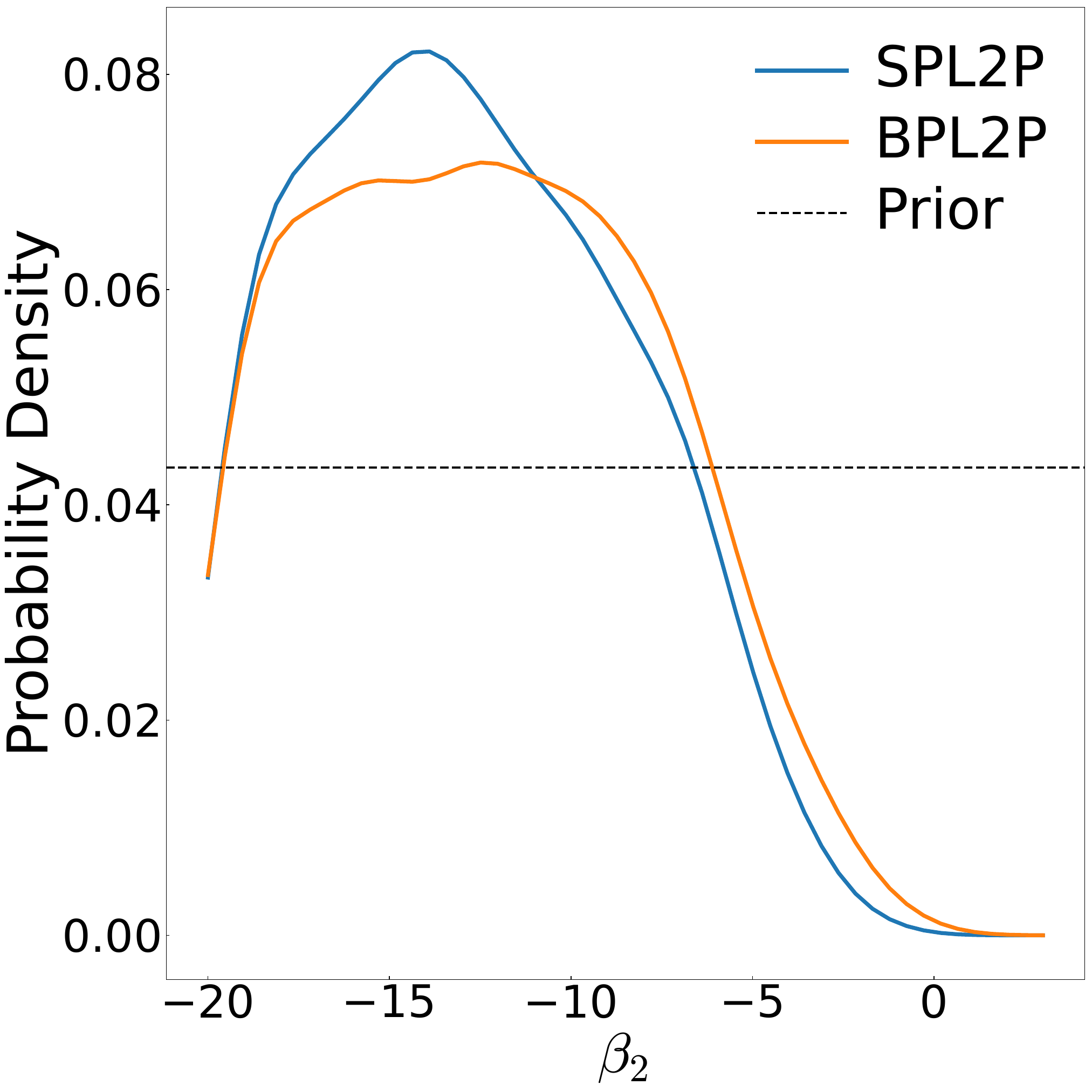}
\caption{\label{fig:m299} Secondary mass percentiles~(\textit{left}) in the range $m_2 \in(m_{2,b1}, m_{2, b2})$ and the corresponding lower bounds on the Carbon-Oxygen reaction rate~(\textit{center}). Note that the bounds of \cite{Tong:2025wpz} are nearly identical to that inferred by \cite{Antonini:2025ilj}. The \textit{right} pannel shows the posterior of $\beta_2$ corresponding to two different $p(m_1)$ models.}
\end{center}
\end{figure*}

Our inferred mass distributions, shown in Figure~\ref{fig:pm1m2}, indicate a break in the powerlaw of mass ratios and a \textit{smooth fall-off} in the merger rate density for $m_{2}\geq m_{2,b1}=40.0^{+9}_{-7}M_{\odot}$. Our model yields a Bayes factor of $191$ relative to the default single-powerlaw mass-ratio model of \cite{LIGOScientific:2025pvj}. As mentioned before, we choose a \texttt{BPL2P} model for the primary-mass distribution to obtain these results. To investigate the model-dependence of our inference, we have explored an alternative model for the primary mass distribution parameterized by a single powerlaw plus two peaks~(SPL2P) and obtained similar results~(see appendix~\ref{sec:appendix-pm1} for the inferred distribution). For the \texttt{SPL2P} primary mass model, our flexible mass-ratio distribution yields a Bayes factor of 353. In both cases, our Bayes factors are three times higher than the gap-inclusive mass-ratio model of \cite{Tong:2025wpz}, who report Bayes factors of 55 and 100 for the \texttt{BPL2P} and \texttt{SPL2P} primary models, respectively. Hence, we conclude that GWTC-4 data prefer a smooth fall-off in merger rate for $m_2\gtrsim 40$ over a sharp drop-off in that region. The fall off in our $m_2$ distribution is consistent with the inference of \cite{Banagiri:2025dmy}, who also constrain mass-based transitions in the mass-ratio distribution without explicitly modeling any gap-like feature.  We demonstrate the prior-dependence of gap recovery in more detail by performing rigorous model-comparison analyses for the mass-distribution in Sec.~\ref{sec:model-comparison}.

Measurement uncertainties indicate that the current detection sample can still allow for a PISN cutoff in $m_2$, albeit at higher masses. We calculate lower-bounds on the gap-onset by computing $m_{2}^{99\%}$ and $m_{2}^{99.9\%}$ for $m_{2,b1}<m_2<m_{2,b2}$ which are shown in Figure~\ref{fig:m299}. We constrain the lower edge of a potential gap in secondary BH masses to be higher than $m_{2}^{99\%}=57^{+17}_{-10}M_{\odot}$, which translates to a Carbon-Oxygen reaction rate constraint~\citep[from the fits of][]{Farmer:2020xne} of $101^{+87}_{23}\mathrm{keV~barns}$ or lower, also shown in Figure~\ref{fig:m299}. We note here that even within the range $m_{2,b1}<m_{2}<m_{2,b2}$, the inference of our percentiles is driven by the data and not the prior, as shown in appendix~\ref{sec:appendix-m299}.

%From these measurements, we conclude that there is no evidence of a PISN mass gap in GWTC-4 data near $40-50M_{\odot}$, which might still exist, at higher masses. The scarcer concentration of events with $m_2>40$ can be explained better by a broken power law than a sharp drop-off, although more data are needed to verify this conclusion with higher confidence. We further identify the location of a plausible sharp PISN gap that is still allowed within the uncertainties of our inferred $m_2$ distributions, lower bounds on whose onset are fully consistent with other measurements of the carbon-oxygen nuclear reaction rate. Our $90\%$ credible interval includes the entire range of values reported by \cite{deBoer:2017ldl} from independent measurements, whereas strongly modeled studies only report marginal overlap~(Figure~\ref{fig:m299}).

From these measurements, we conclude that there is no evidence of a PISN mass gap in GWTC-4 data near $40-50M_{\odot}$, which might still exist at higher masses. The scarcer concentration of events with $m_2\gtrsim40M_{\odot}$ can be explained better by a broken power law than a sharp drop-off. Note that this conclusion is driven by the shape of the $m_2$ distribution, which in turn is determined by the posterior of $\beta_2$. As we show in Figure~\ref{fig:m299}, and later on in Sec.~\ref{sec:model-comparison} with broader priors, $\beta_2$ has significant posterior support for values that are not consistent with a sharp drop-off and is bounded from below. The percentiles of the $m_2$ distribution further inform on the lower edge of a plausible mass-gap that is still allowed within the measurement uncertainties of the current detection sample. In other words, our $\beta_2$ posterior and the corresponding shape of the $m_2$ distribution indicate no evidence of a sharp upper mass gap in $m_2$ given GWTC-4.

However, more data are needed to verify this conclusion with higher confidence. We further identify the location of a plausible sharp PISN gap that is still allowed within the uncertainties of our inferred $m_2$ distributions, lower bounds on whose onset are found to be $\simeq 60M_{\odot}$, which is fully consistent with other measurements of the carbon-oxygen nuclear reaction rate. Our $90\%$ credible interval includes the entire range of values reported by \cite{deBoer:2017ldl} from independent measurements, whereas strongly modeled studies only report marginal overlap~(Figure~\ref{fig:m299}).

% \begin{itemize}
%     \item{Bayes factor table}
%     \item{Comparison with BGP}
% \end{itemize}

\section{Spin and mass-ratio transition: Alternatives to the hierarchical merger scenario}
\label{sec:spin-dist}
Regardless of an $m_2$ gap, other observational signatures of the PISN cut-off near $\simeq 40M_{\odot}$ have been reported in GWTC-4 that have been claimed as evidence in favour of a sub-population of (2G+1G) hierarchical mergers above that mass range. Despite the evidence in favour of a smooth fall-off in $m_2$ over a sharp gap, the findings presented in the previous section do not rule out a sub-population above $m_1\geq 40M_{\odot}$ with a distinct effective spin distribution and a mass-ratio distribution consistent with the 2G+1G scenario. To address this question, we model the spin transition predicted by \cite{Rodriguez:2019huv, Antonini:2024het}, and confirmed by \cite{Antonini:2025zzw} in GWTC-3, and by \citep{Antonini:2025ilj, Tong:2025wpz} in GWTC-4.
\begin{figure*}[htt]
\begin{center}
\includegraphics[width=0.32\textwidth]{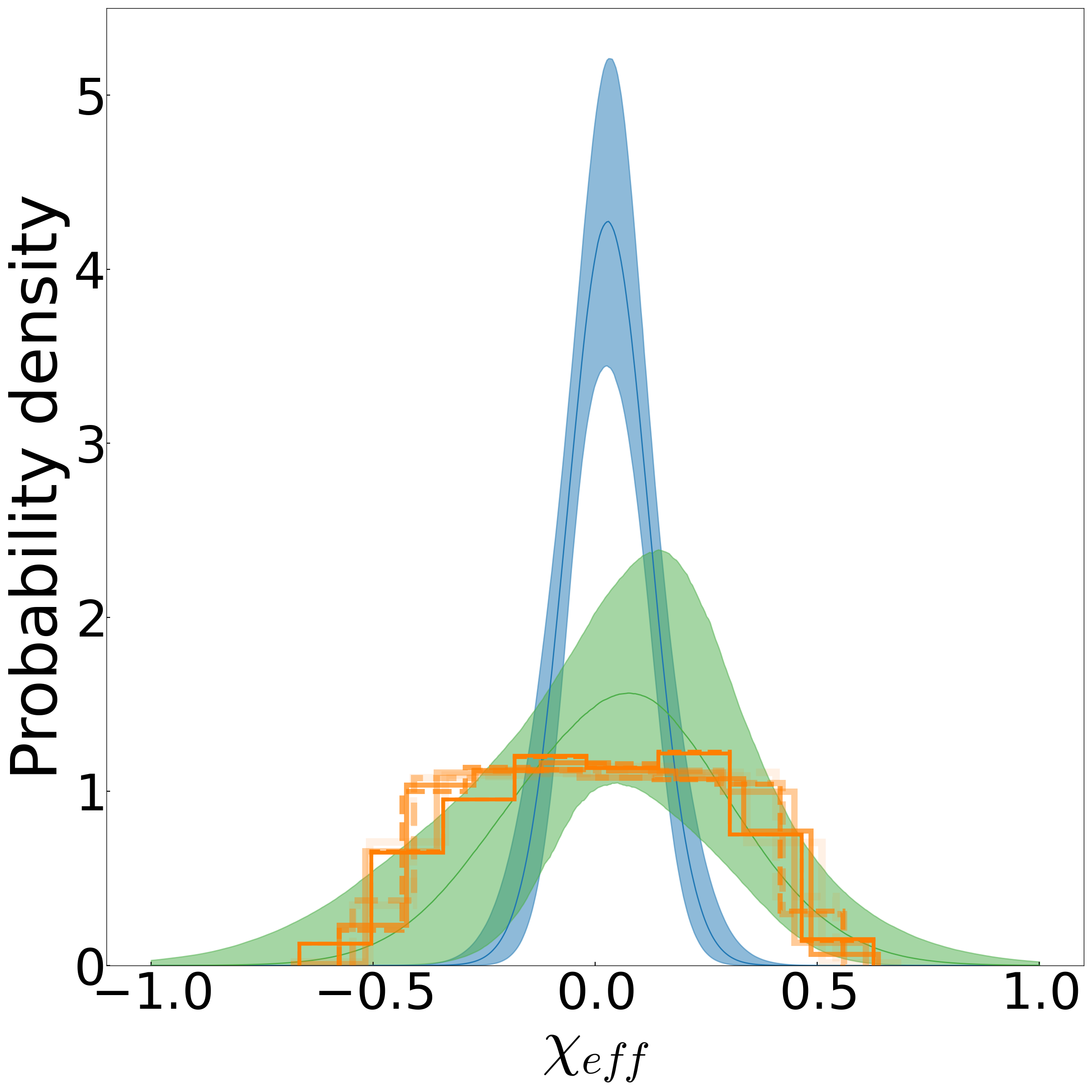}
\includegraphics[width=0.32\textwidth]{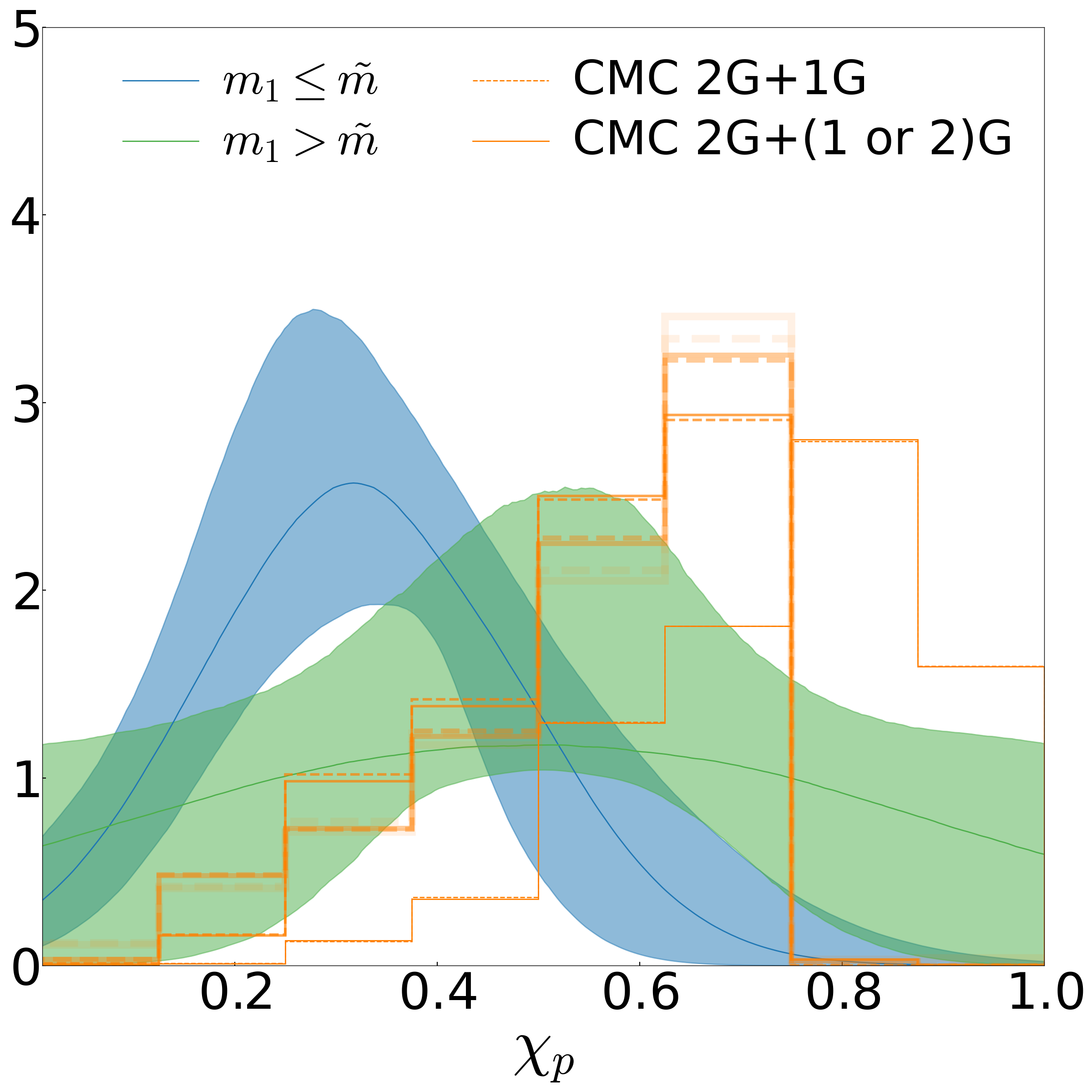}
\includegraphics[width=0.32\textwidth]{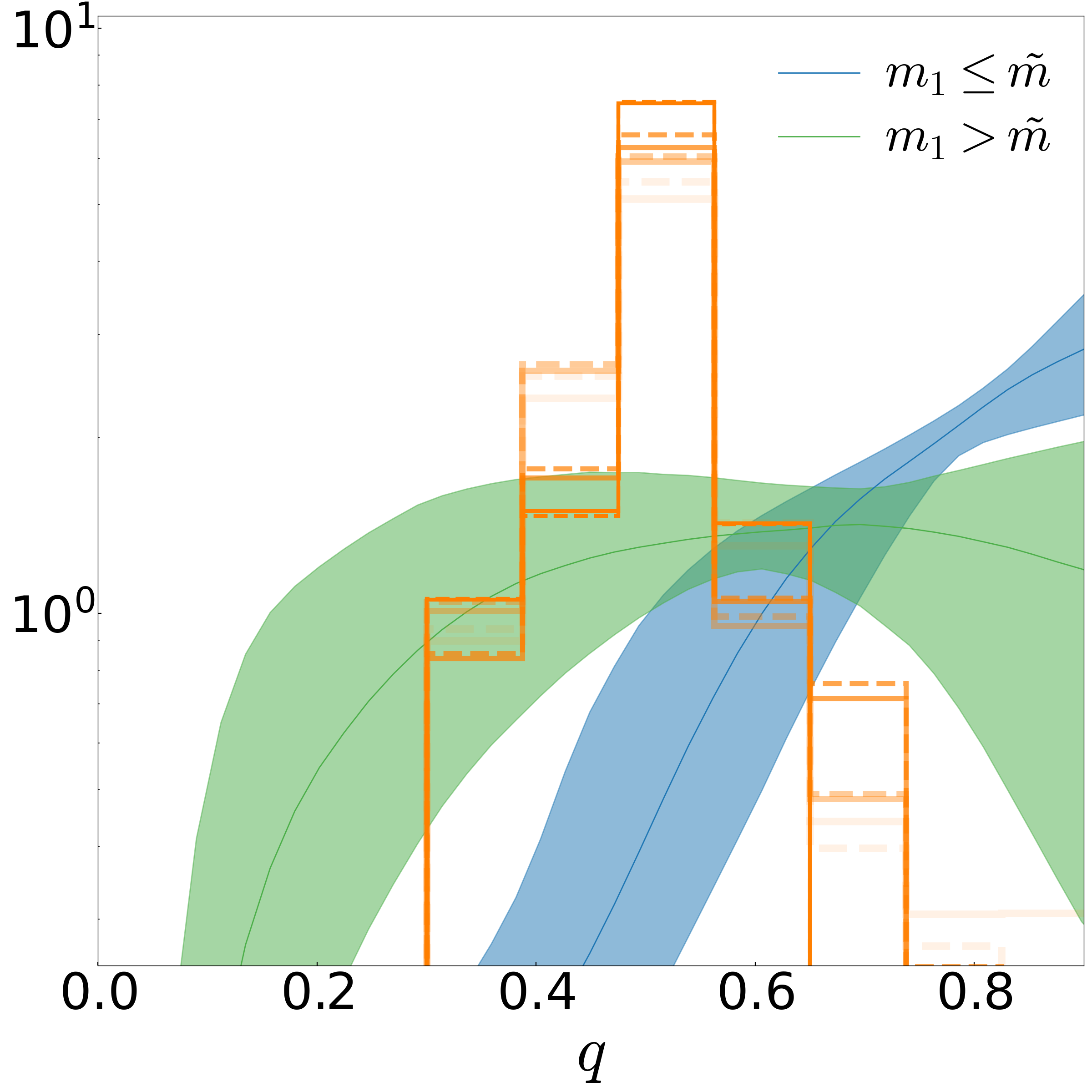}
\caption{\label{fig:trans} Transition of mass-ratio, effective aligned and effective precision-spin distributions at $\tilde{m}=43_{-5}^{+11}M_{\odot}$. The CMC simulations were carried out for different values of the BBH birth spin which are indicated by the shade of orange. The faintest to brightest lines correspond to birth spins of $[0, 0.1, 0.2, 0.5]$.}
\end{center}
\end{figure*}
% /home/christian.adamcewicz/projects/mass_spin_copula/transition/spin-transition-models.py
%/home/christian.adamcewicz/projects/o4/o4b-populations/copulas/gwpop_config_mass_1_chi_eff_copula_gwtc5.ini

To constrain any transitions of mass-ratio, effective aligned~$(\chi_{eff})$ and effective precessing~$(\chi_p)$ spin distributions across various mass ranges, we use the following models:
\begin{equation}
    p(\chi_{eff}|m_1) = \begin{cases}
        \mathcal{TN}_{\chi}(\chi_{eff},\mu_{1,\chi_{eff}}, \sigma_{1,\chi_{eff}}, -1,1), & m_1 \leq \tilde{m},\\
        \mathcal{TN}_{\chi}(\chi_{eff},\mu_{2,\chi_{eff}}, \sigma_{2,\chi_{eff}}, -1,1), & m_1 > \tilde{m}
    \end{cases},\label{eq:chieff-dist}
\end{equation}
and
\begin{equation}
    p(\chi_{p}|m_1) = \begin{cases}
        \mathcal{TN}_{\chi}(\chi_{p},\mu_{1,\chi_{p}}, \sigma_{1,\chi_{p}}, -1,1), & m_1 \leq \tilde{m},\\
        \mathcal{TN}_{\chi}(\chi_{p},\mu_{2,\chi_{p}}, \sigma_{2,\chi_{p}}, 0,1), & m_1 > \tilde{m}\label{eq:chiep-dist}
    \end{cases},
\end{equation}
in conjunction with our flexible mass-ratio distribution. See appendix~\ref{secc:appendix-dist} for exact functional forms. We also consider an alternative model where the $\chi_{eff}$ and $\chi_p$ distributions below and above $m_1=\tilde{m}$ are allowed to be correlated,  and obtain similar results, although the models in Eqs.~\eqref{eq:chieff-dist}, and \eqref{eq:chiep-dist}, are preferred by the data over the correlated one, likely due to an Occam's penalty. In conjunction with our mass-ratio distribution, these spin models can be used to shed light on the astrophysical origins of the sub-population of BBHs at $m_1\geq \tilde{m}$.

% \cite{Antonini:2024het, Rodriguez:2019huv}, and reported by \cite{Antonini:2025zzw} in GWTC-3, and by \citep{Antonini:2025ilj, Tong:2025wpz} in GWTC-4.
\begin{figure*}[htt]
\begin{center}
\includegraphics[width=0.32\textwidth]{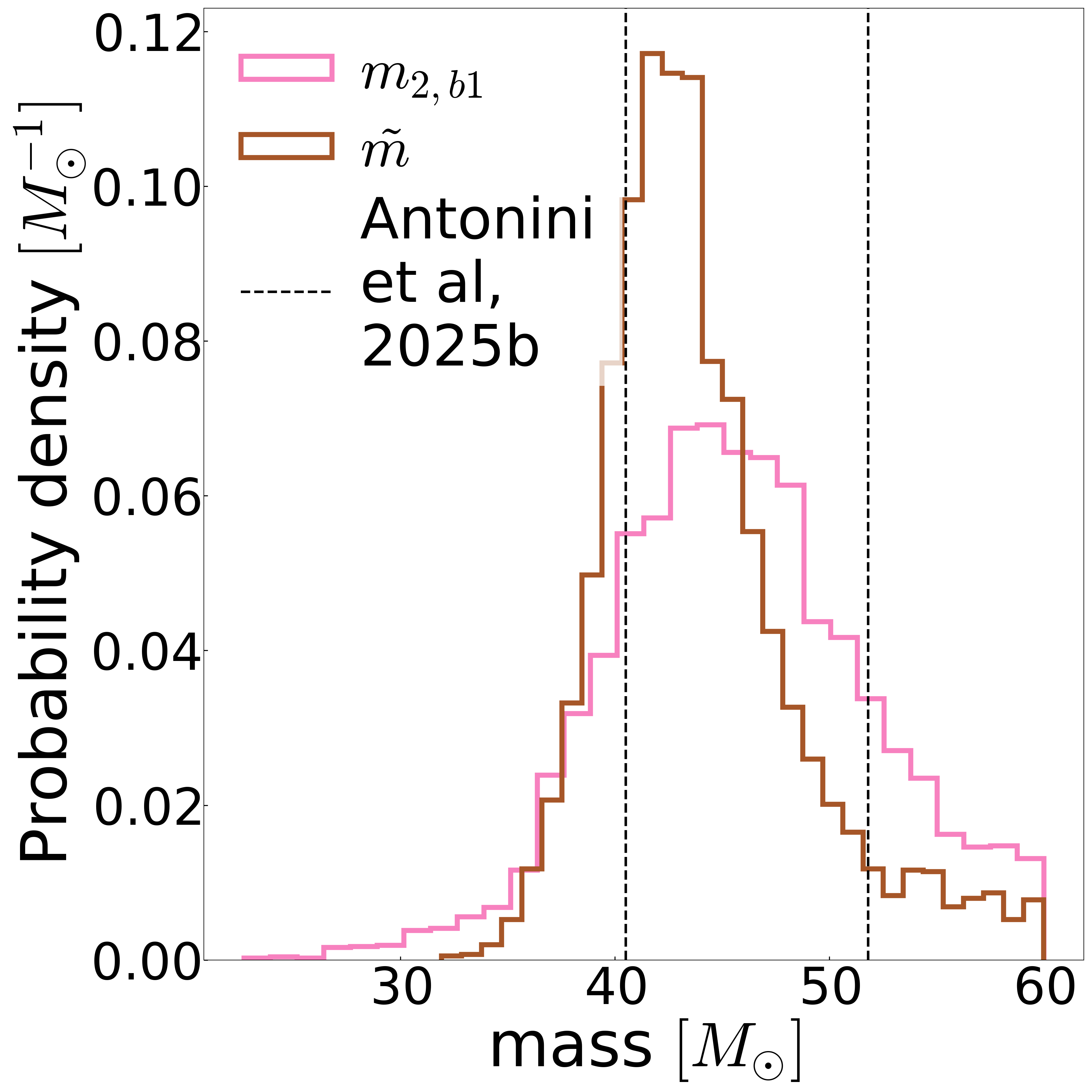}
\includegraphics[width=0.32\textwidth]{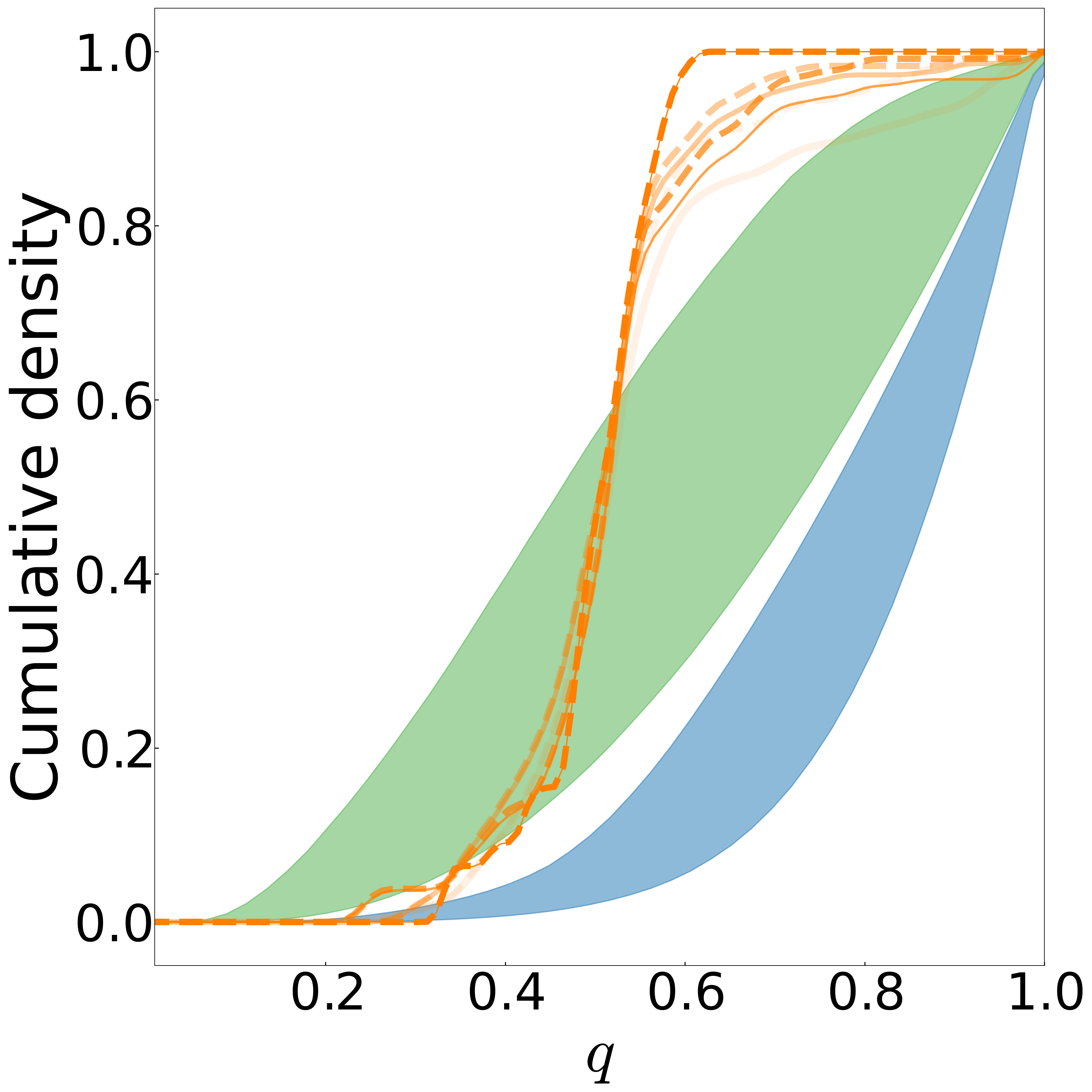}
\includegraphics[width=0.32\textwidth]{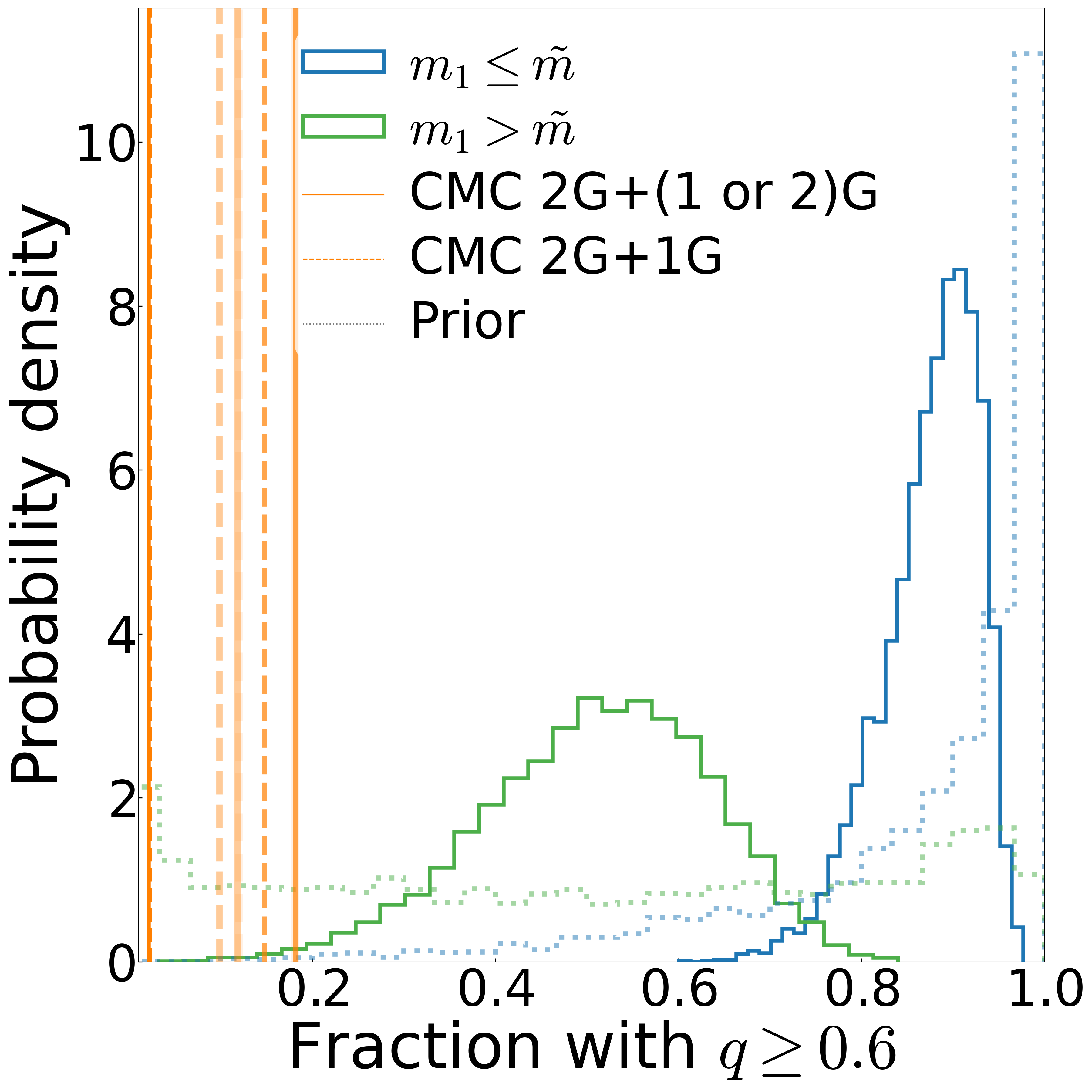}
\caption{\label{fig:qfrac} Metrics from the spin transition study: the transition mass compared to the break location in terms of secondary mass of the mass-ratio distribution~(left), cumulative density function~(center), and fraction of events with $q\geq 0.6$~(right), below and above the spin-transition mass. As in Figure~\ref{fig:trans}, for the CMC simulations, different shades of orange indicate different values of the BH birth spin.}
\end{center}
\end{figure*}

We infer these spin distributions simultaneously with our flexible mass-models to characterize the sub-population of BBHs above the spin-transition mass~$(\tilde{m})$, which we allow to be different from $m_{2,b1}$. We compare our post-transition sub-population with simulations of BBH formation in globular clusters~\citep{Rodriguez:2019huv} using the publicly available code Cluster Monte Carlo~\citep[CMC,][]{Pattabiraman:2012ti}. The simulations were taken from the public data release by \cite{michael_zevin_2020_4277620}. We also compare with the latest CMC catalog in appendix~\ref{sec:appendix-cmc} and obtain similar results. As shown in figure~\ref{fig:trans}, we find above the transition mass which we constrain to be $43_{-5}^{+11}M_{\odot}$, a broadening of the $\chi_{eff}$ and $\chi_p$ distributions, in agreement with \cite{Tong:2025wpz, Antonini:2025ilj}, and the theoretical predictions of dynamical simulations~\citep{Rodriguez:2019huv, Antonini:2024het}. However, the corresponding mass-ratio distribution has a significantly higher preference for symmetric mass systems than expected for (2G+1G) hierarchical mergers. Furthermore, the theoretical predictions are dominated by 2G+1G systems, and contributions from 2G+2G mergers cannot resolve the inconsistency with our inferred distributions.

In Figure~\ref{fig:qfrac}, we show the posterior for the fraction of events with mass-ratio greater than 0.6 below and above the spin-transition mass~$(\tilde{m})$, with the post-transition sub-population having a $q\in(0.6-1)$ fraction of $52^{+18}_{-23}\%$. The mass-ratio distributions employed by \cite{Antonini:2025ilj} enforce a single power-law independent of primary mass and are hence incapable of inferring a change in the mass-ratio distribution above the spin-transition mass. The models of \cite{Tong:2025wpz} impose strong priors on the mass-ratio distribution, which can drive their inference towards low mass-ratios in the $m_1>\tilde{m}$ sub-population. To show this, we perform additional model comparison analyses between the flexible and gap-enforcing priors, now allowing for spin-transition, in section~\ref{sec:model-comparison}. 

%As seen in figure ()in the context of our analogous gap enforced priors in figure ().

The results presented in this section indicate that the GWTC-4 sample of BBHs is not fully consistent with the sub-population above $m_1\sim 40M_{\odot}$ being solely comprised by (2G+1G) hierarchically merging systems. This inconsistency is not resolved through the contributions of 2G+2G systems which are not abundant enough to increase support in the theoretical $q$ distributions for $q>0.6$. However, the existence of spin-transition remains robust. The post-transition sub-population is potentially also difficult to explain by systems originating only from isolated stellar binaries, due to the preference for higher spin-precession and weaker spin-alignment. We conclude that alternatives to the hypothesis that the $m_1>\sim 40M_{\odot}$ sub-population is predominantly hierarchical in origin merit consideration given GWTC-4 data. Our conclusions are in principle, consistent with the findings of \cite{Banagiri:2025dmy}, who characterize the distributions of component spin magnitudes and tilts in this high-mass sub-population, and find features (such as support for low primary spins, high secondary spins and no transition in the spin tilt distributions) that are also difficult to explain exclusively with the (2G+1G) hierarchical merger scenario. We note, however, that unlike the spin transition study of \cite{Banagiri:2025dmy}, we allow the transition mass to be different for effective spins and mass ratios and infer a common transition scale a posteriori.

\section{Model comparison: Enforcing a gap with strong priors}
\begin{figure*}[htt]
    \centering
    \includegraphics[width=0.98\linewidth]{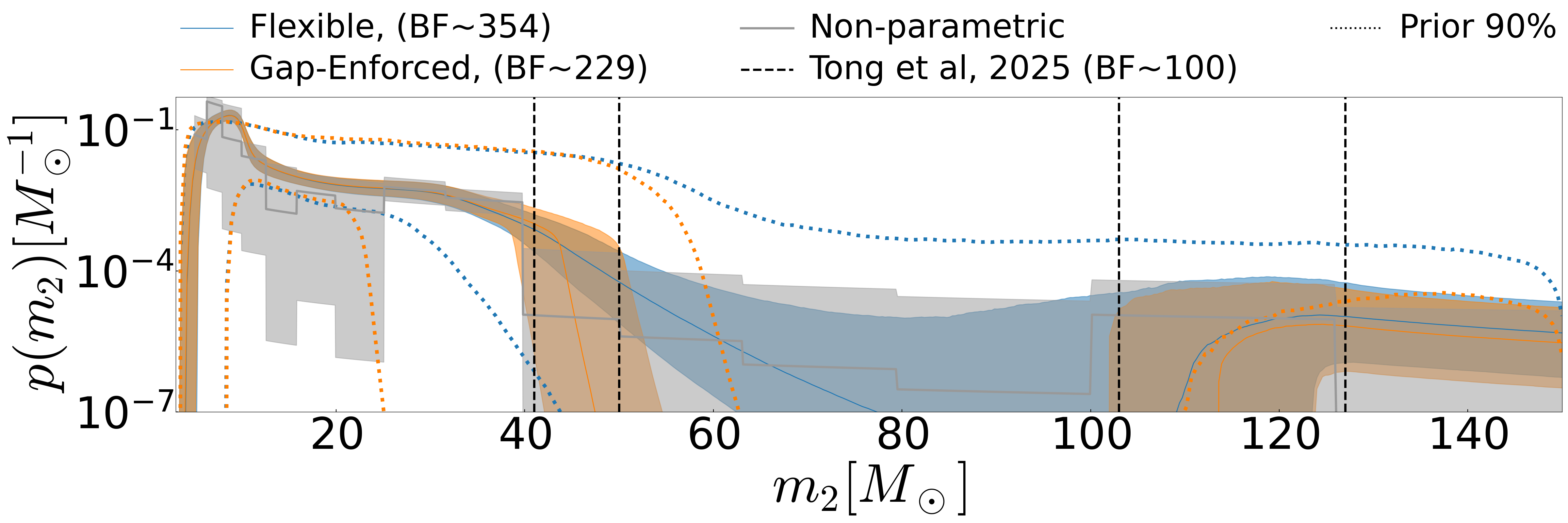}
    \includegraphics[width=0.98\linewidth]{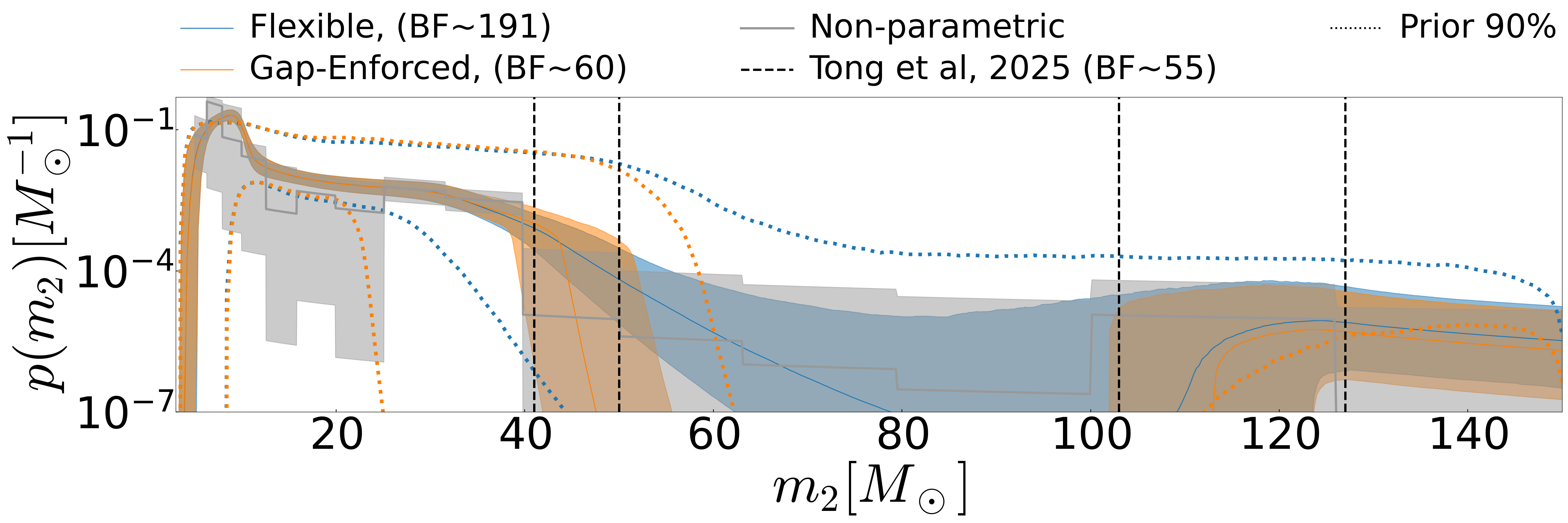}
    \caption{Comparison of the gap enforced inference with the flexible one and non-parametric results. The top panel uses \texttt{BPL2P} for the primary mass model and the bottom one \texttt{SPL2P}. The non-parametric (binned Gaussian process) results for GWTC-4 were reported in \cite{LIGOScientific:2025pvj} and are obtained from their public data release: \cite{ligo_scientific_collaboration_2025_16911563}. The black dashed lines demarcate the lower and upper edge of the PISN mass-gap inferred by \cite{Tong:2025wpz} using their gap-inclusive model. The coloured dotted lines demarcate the $90\%$ credible intervals of the prior predictive distribution corresponding to each prior choice.}
    \label{fig:model-comparison}
\end{figure*}
\label{sec:model-comparison}
\begin{table}[h]
\centering
\begin{minipage}{0.48\textwidth}
\centering
\begin{tabular}{ccc}
\hline
\hline
Model  & $p(q)$  & BF\\
\hline
Default &  single-powerlaw & 1 \\
Flexible & Eq.~\eqref{eq:q-dist}, Narrow $\beta_2$ prior&  354\\
Broad & Eq.~\eqref{eq:q-dist}, Broad $\beta_2$ prior&  340\\
Gap enforced & Eq.~\eqref{eq:q-dist}, Restricted priors   & 229\\
Tong et al 2025 &  Single Powerlaw + Step/Notch & 100\\
\hline
\end{tabular}
\end{minipage}
\begin{minipage}{0.48\textwidth}
\centering
\begin{tabular}{ccc}
\hline
\hline
Model  & $p(q)$  & BF\\
\hline
Default &  single-powerlaw & 1 \\
Flexible & Eq.~\eqref{eq:q-dist}, Narrow $\mathbf{\beta_2}$ prior &  191\\
Broad &  Eq.~\eqref{eq:q-dist}, Broad $\beta_2$ prior&  95\\
Gap enforced & Eq.~\eqref{eq:q-dist}, Restricted priors   & 60\\
Tong et al 2025 &  Single Powerlaw + Step/Notch & 55\\
\hline
\end{tabular}
\end{minipage}
\caption{Bayes factors for the mass-distribution inference, for two different primary-mass models: \texttt{SPL2P} \textit{(left)} and \texttt{BPL2P} \textit{(right)}. }
\label{table:bayes-fac}
\end{table}

Our flexible model can fit for a gap but does not enforce one. The results reported by \cite{Tong:2025wpz} correspond to mass-ratio models that comprise either a notch filter or a box function, and incorporate a prior on the gap width to be greater than $20M_{\odot}$. To assess the role of such priors, we can enforce a similar gap in our model as well, by imposing strong assumptions on the relevant parameters such as $\beta_2\in(-140,-70),~m_{2,b2}\in(100, 200)$. Under this assumption, we indeed recover a gap, much like the one reported in~\cite{Tong:2025wpz}, but the corresponding Bayes factor is 3 times smaller than that of the flexible inference. We also analyze the data with a broad prior on $\beta_2$ encompassing both ranges: $\beta_2\in(-200,3)$. We further explore the dependence on primary-mass models. Similar to \cite{Tong:2025wpz}, we perform model selection among different mass-ratio distributions and two alternatives for the primary mass distribution, namely the \texttt{BPL2P}, and $\texttt{SPL2P}$.

In Table~\ref{table:bayes-fac} we compare the Bayes factors of our flexible and gap-enforced inferences with those of the corresponding gap-inclusive model of \cite{Tong:2025wpz}. In Figure~\ref{fig:model-comparison}, we also compare our inferences~(i.e. the $90\%$ credible intervals of both the posterior and prior predictive distributions) with a non-parametric reconstruction of the secondary mass-distribution using binned Gaussian Processes obtained from the same dataset~\citep{Ray:2023upk, Ray:2024hos, Mohite:2022pui, KAGRA:2021duu, LIGOScientific:2025pvj}, which were taken from the public data-release by the LVK at \cite{ligo_scientific_collaboration_2025_16911563}.

It can be seen that the gap-enforced inference is likely prior-driven and that the non-parametric constraints are fully consistent with a smooth fall off as recovered by our flexible model. In conjunction with the reported Bayes factors, these results indicate that the recovery of a sharp gap in the $m_2$ distribution from GWTC-4 can, in general, be prior-driven and that the data prefer a smoother fall-off in the merger rate density at $m_2\gtrsim 40M_{\odot}$. In particular, it can be seen that the gap-enforced case is disfavoured by the data, not due to an increase in prior volume but because the gap-enforcing prior drives the inference towards a poorer fit. Furthermore, the inference with the broader prior clearly demonstrates that posterior support declines towards more negative values of $\beta_2$, particularly in regions of the parameter space that are consistent with a sharp gap. Note, however, that the Bayes factor corresponding to the broad prior is slightly smaller than that of the narrow one, which is expected given the much larger prior volume.
\begin{table}[h]
\centering
\begin{tabular}{ccccc}
\hline
\hline
Model & $p(m_1)$ & $p(q)$ & $p(\chi_{eff}, \chi_p)$  & BF\\
\hline
Default & \texttt{BPL2P} & single-powerlaw & Single Gaussian (mass independent)& 1 \\
Default+transition & \texttt{BPL2P} & single-powerlaw & Eqs.~\eqref{eq:chieff-dist},\eqref{eq:chiep-dist}(mass-dependent transition)& 281\\ 
Flexible+transition & \texttt{BPL2P} & Eq.~\eqref{eq:q-dist} & Eqs.~\eqref{eq:chieff-dist},\eqref{eq:chiep-dist}(mass-dependent transition) & 1730\\
Gap enforced+transition & \texttt{BPL2P} & Eq.~\eqref{eq:q-dist},(restricted priors)   &  Eqs.~\eqref{eq:chieff-dist},\eqref{eq:chiep-dist}(mass-dependent transition) & 394\\
\hline
\end{tabular}
\caption{Bayes factors including spin-transition models. }
\label{table:bayes-fac-spin}
\end{table}
\begin{figure*}[htt]
\begin{center}
\includegraphics[width=0.31\textwidth]{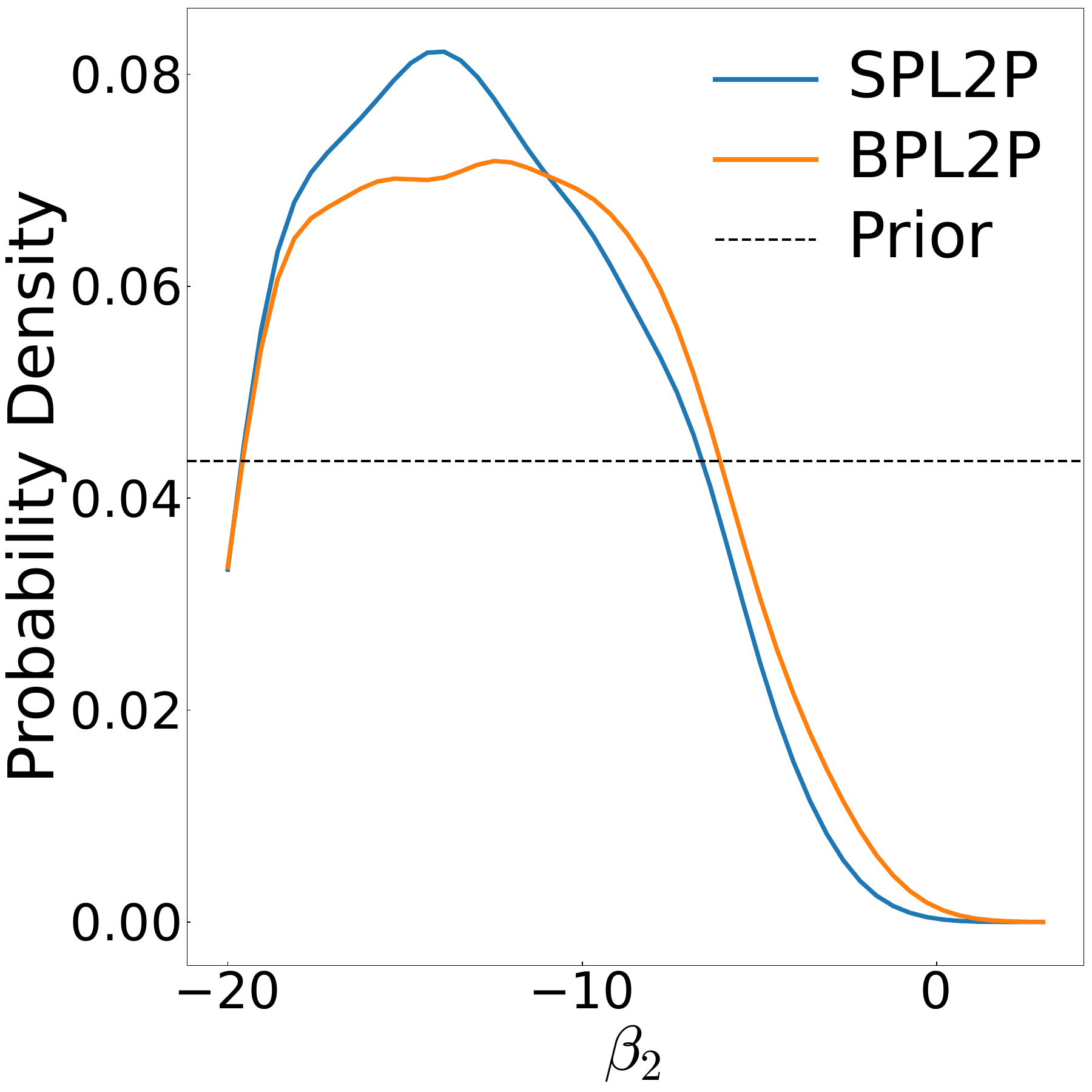}
\includegraphics[width=0.31\textwidth]{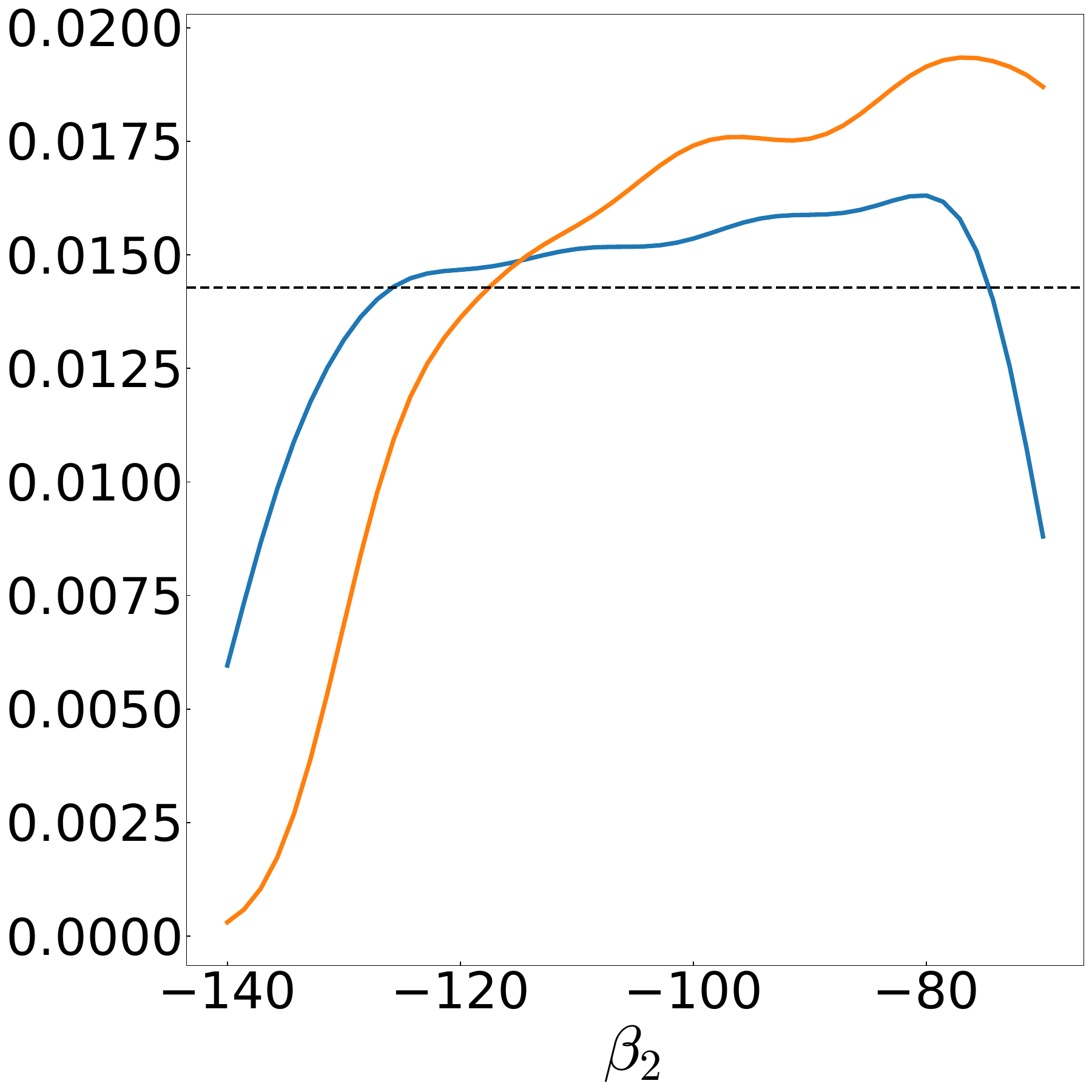}
\includegraphics[width=0.31\textwidth]{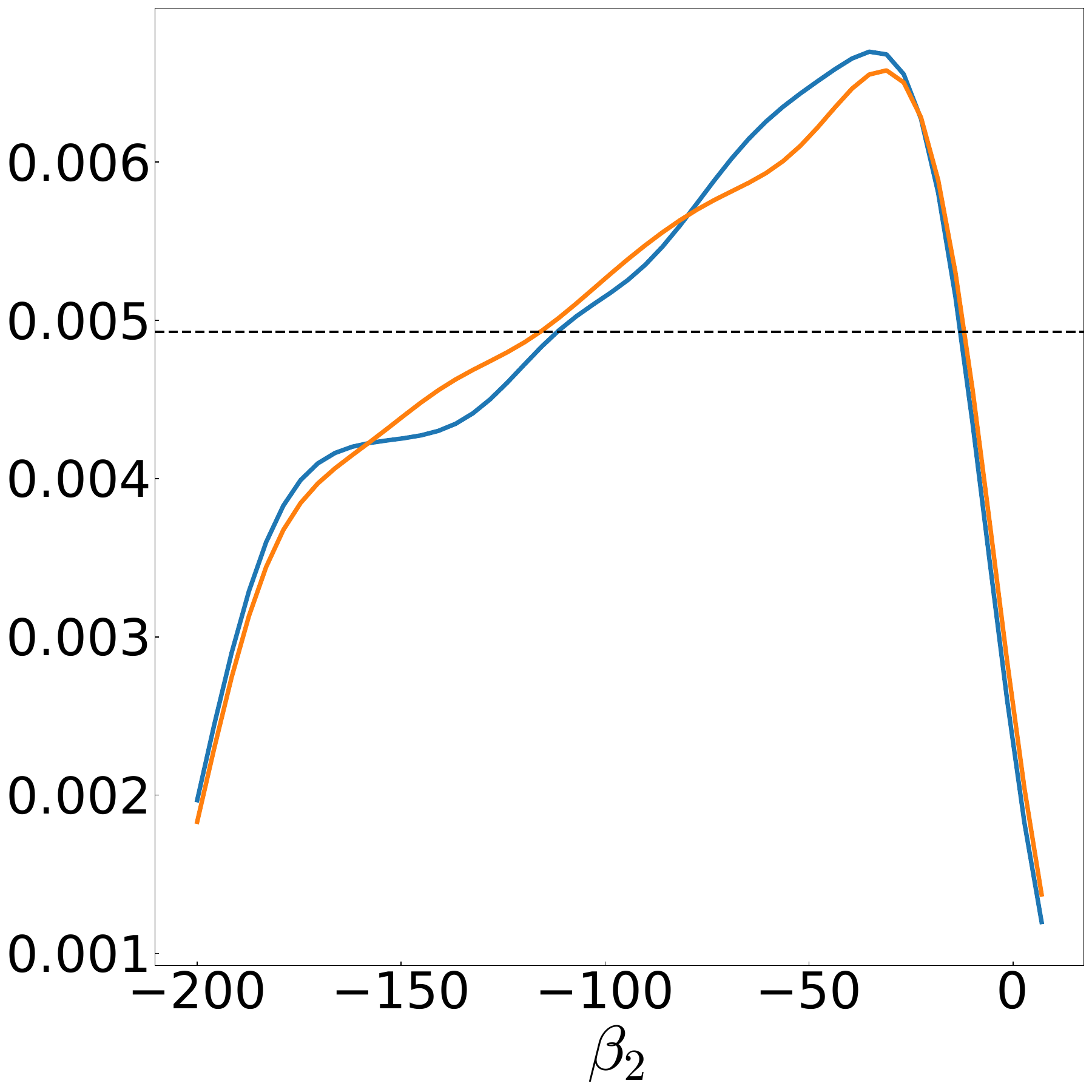}
\caption{\label{fig:beta-post} Posterior distribution of $\beta_2$ for the two $p(m_1)$ models, corresponding to the narrow~(\textit{left}), gap-enforced~(\textit{center}) and broad~(\textit{right}) priors respectively.}
\label{fig:beta-post}
\end{center}
\end{figure*}

\begin{figure*}[htt]
\begin{center}
\includegraphics[width=0.32\textwidth]{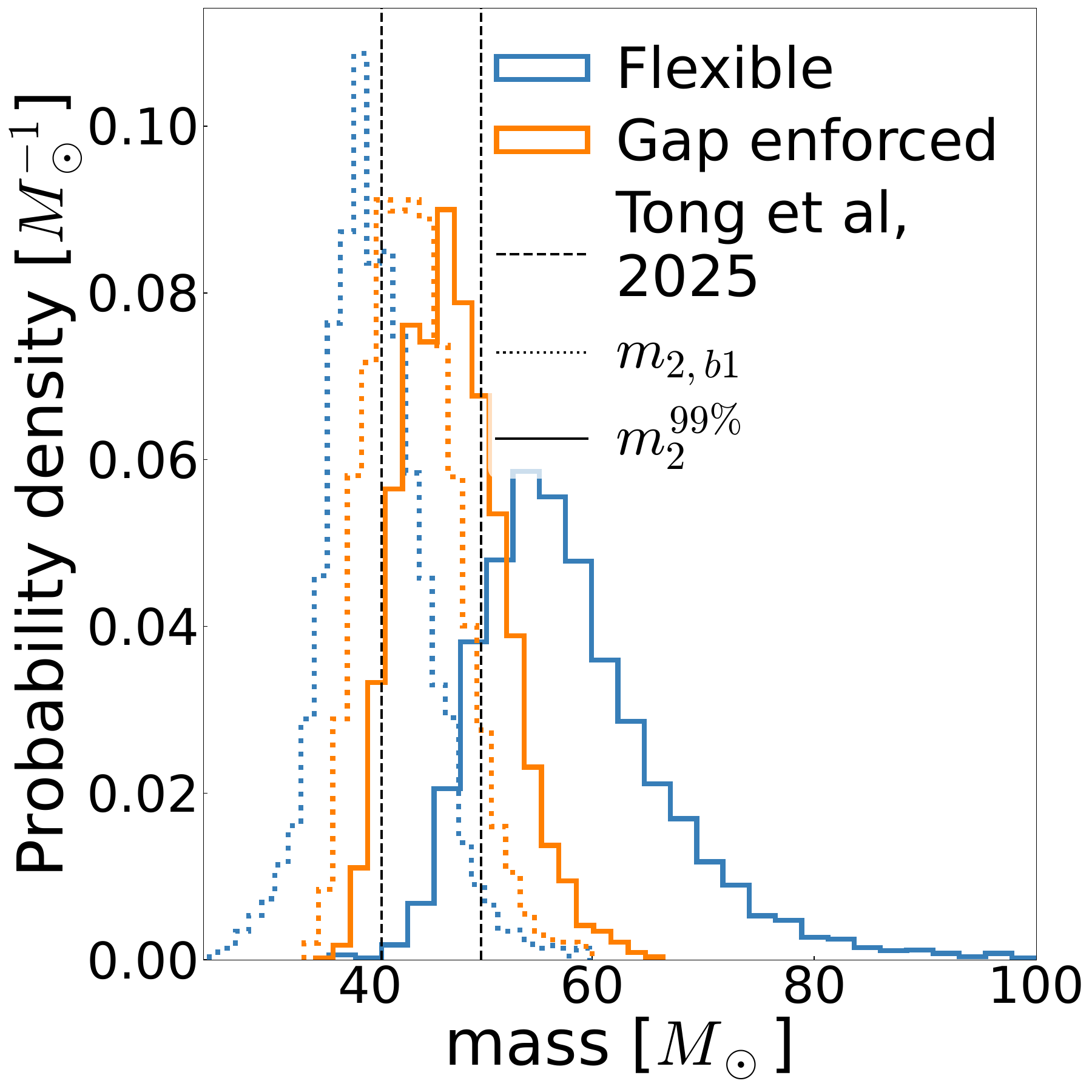}
\includegraphics[width=0.32\textwidth]{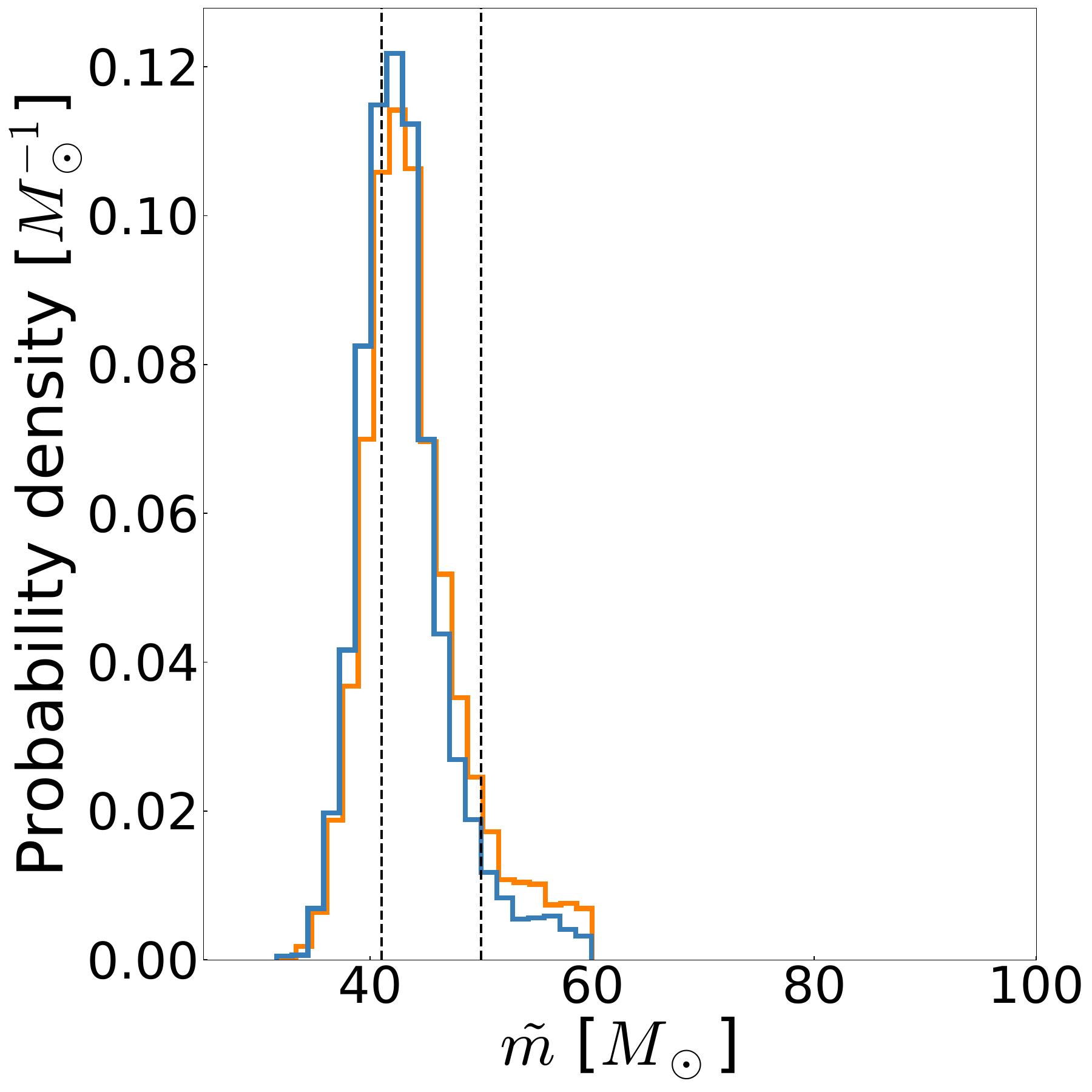}
\includegraphics[width=0.32\textwidth]{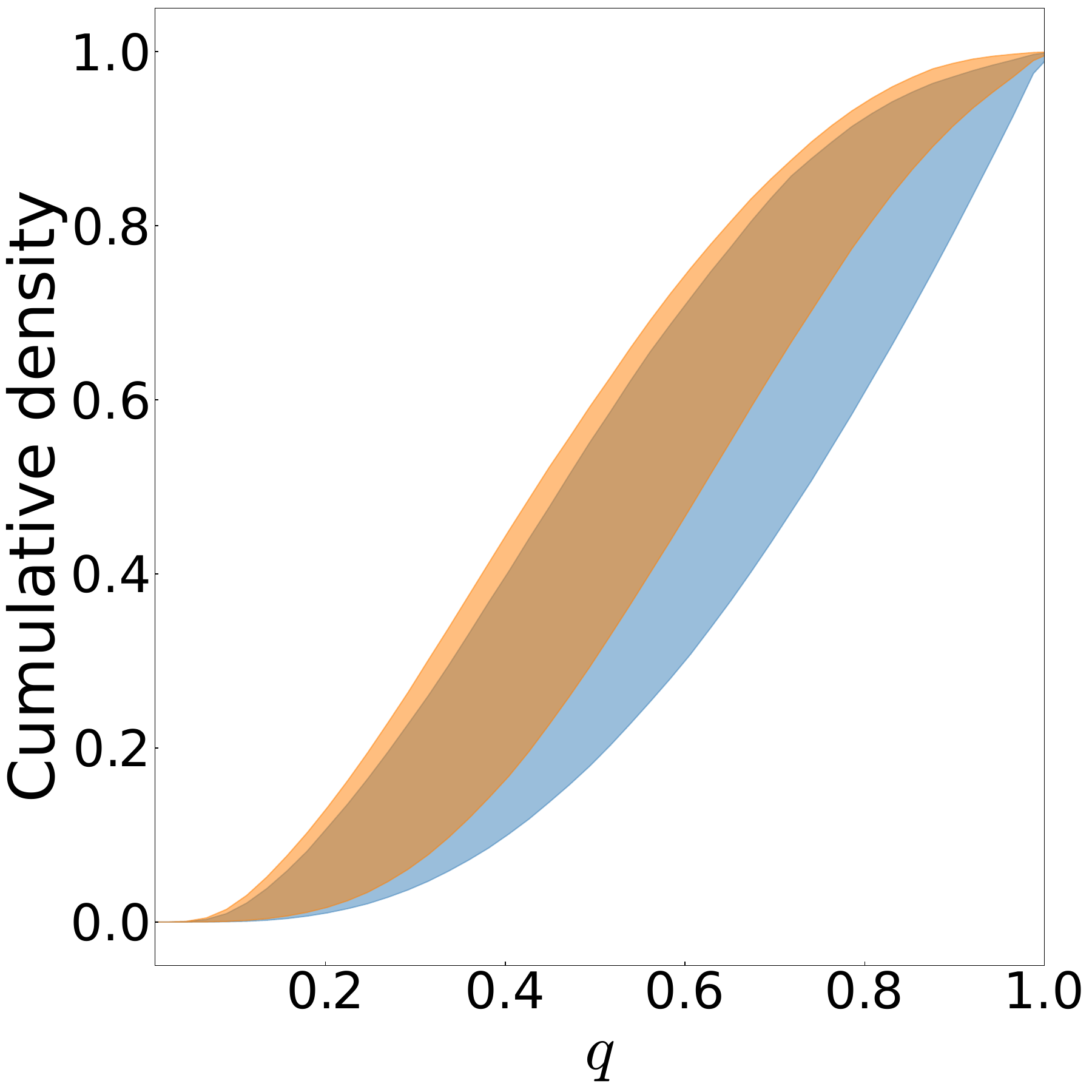}
\caption{\label{fig:bad-beta} Comparison with gap enforced prior for metrics that can identify an underlying PISN cut-off. The dashed lines in the middle panel denote the constraints of \cite{Antonini:2025ilj} and the cumulative densities on the right panel are for the $m_1>\tilde{m}$ sub-population.}
\label{fig:bad-beta-met}
\end{center}
\end{figure*}
Next, we repeat this study, including spin-transition models, to demonstrate the impact of restrictive priors on the inferred mass-ratio distributions of the sub-population above the spin-transition mass. Figure~\ref{fig:bad-beta-met} shows that a gap-enforcing prior can drive this inference towards being more consistent with the prediction that the post-transition subpopulation might be hierarchical in origin. However, the Bayes factors in table~\ref{table:bayes-fac-spin} show that the data prefer the flexible inference over the gap-enforced one by a factor of $~4$. In other words, there is evidence against strong mass asymmetry in the post-spin-transition high mass sub-population, making it difficult to explain with the 2G+1G hierarchical merger scenario. To further validate this conclusion, we perform additional tests for model-systemics in our inference, including posterior predictive checks~\citep{Fishbach:2019ckx, Callister:2023tgi, Miller:2024sui}, which are summarized in appendix~\ref{sec:appendix-ppd}. 

\section{Astrophysical implications}
\begin{figure*}[htt]
\begin{center}
\includegraphics[width=0.32\textwidth]{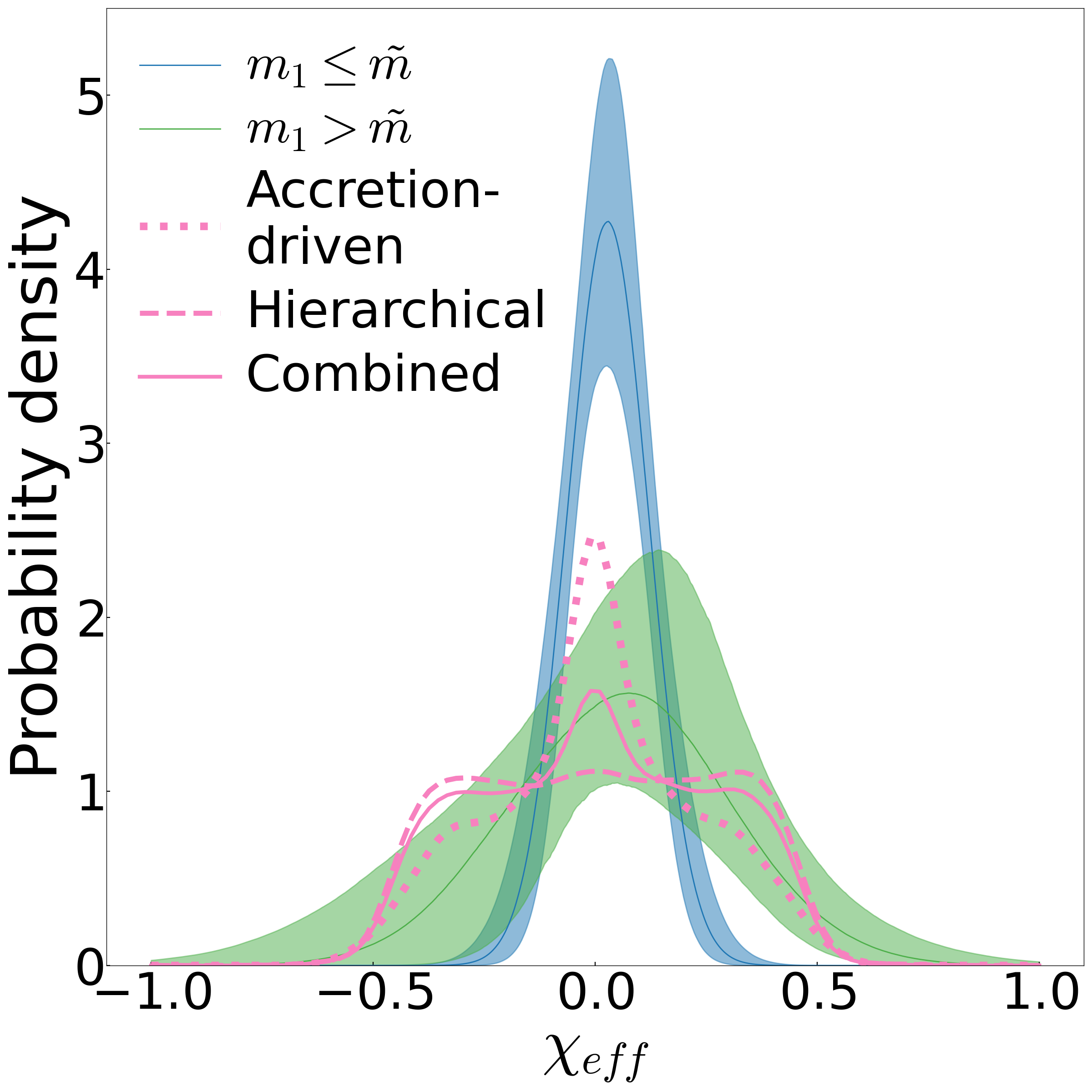}
\includegraphics[width=0.32\textwidth]{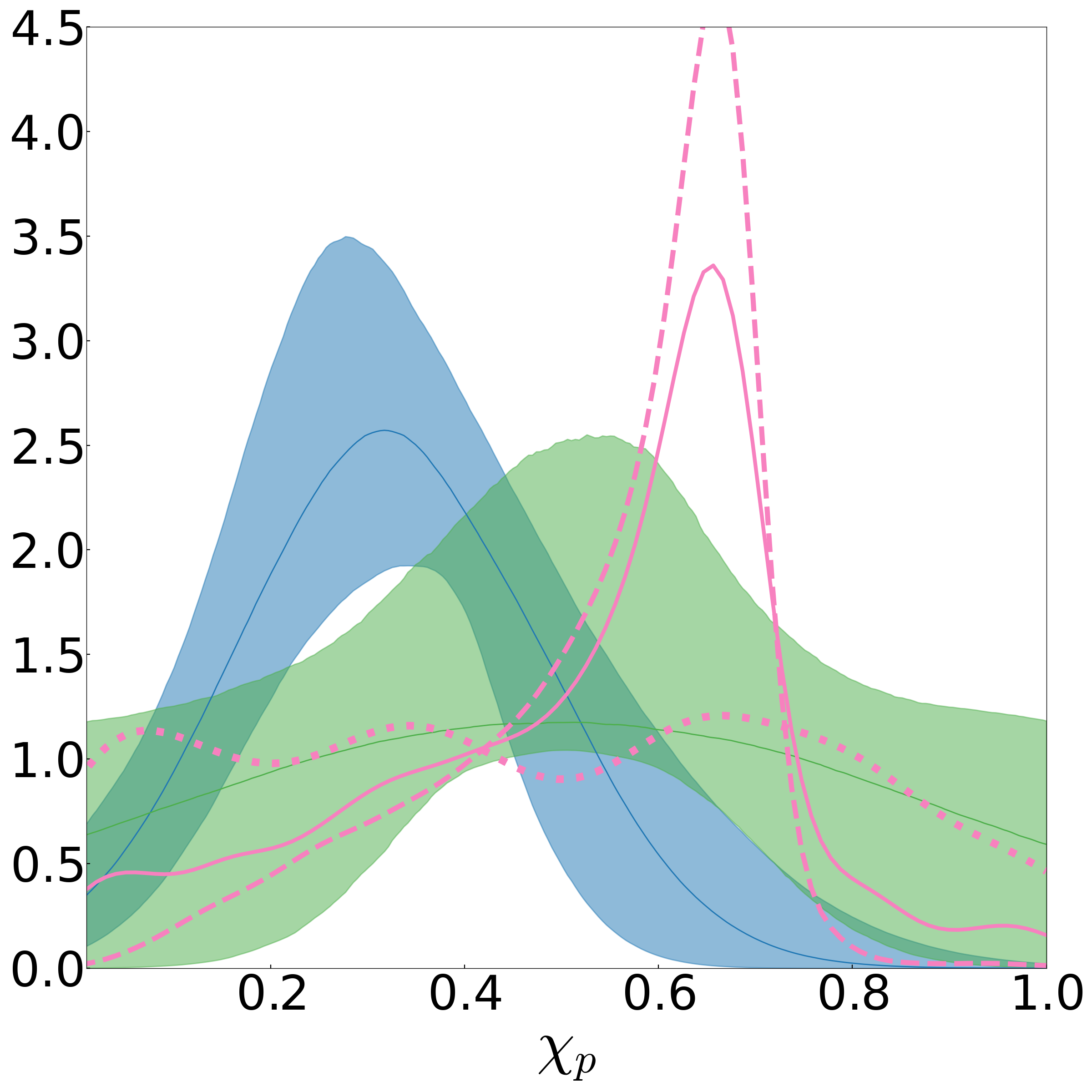}
\includegraphics[width=0.32\textwidth]{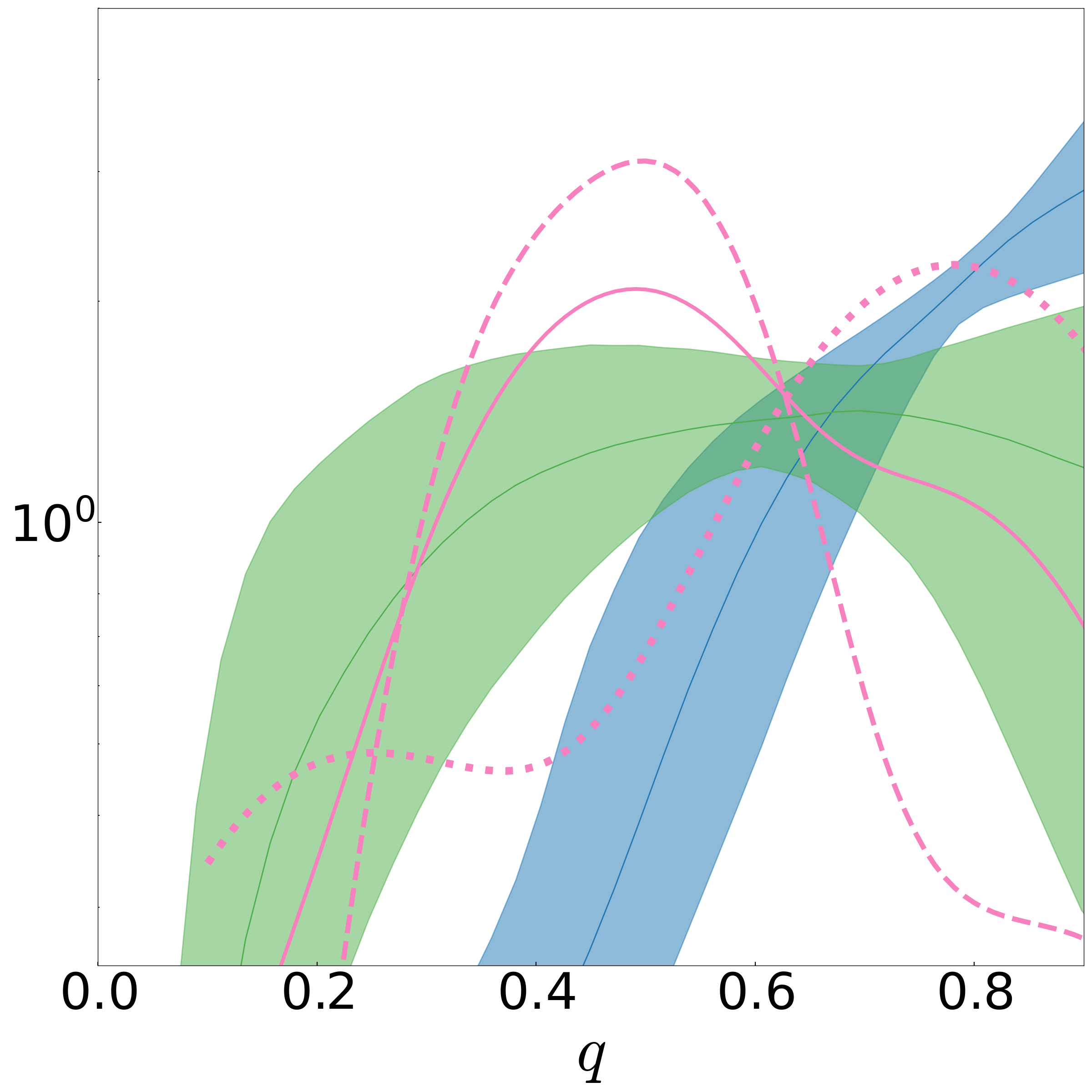}
\caption{\label{fig:trans-acc} Comparing the simulations of \cite{Kiroglu:2025vqy} with our inferred distributions above the spin-transition mass. The pink lines correspond to the models of \cite{Kiroglu:2025vqy} that assume $50\%$ accretion efficency post BH-star collisions\textsuperscript{\ref{fn:fulya}}, and only comprise systems with $m_1\geq 40M_{\odot}$. The relative abundance of the hierarchical and accretion-driven channels is intrinsic to the dynamical simulation and its underlying physical assumptions.}
\end{center}
\end{figure*}
\label{sec:conclusion}
In this letter, we have shown that the GWTC-4 detection sample of BBHs does not exhibit clear observational signatures of a PISN cut-off in the BH mass-spectrum near $40-50M_{\odot}$. Any potential gap still allowed within measurement uncertainties may onset at masses $(57^{+17}_{-10}M_{\odot})$ or higher, indicating an S factor of $^{12}\mathrm{C}(\alpha, \gamma)^{16}O$ at 300 kev of $101^{+87}_{-23}\mathrm{keV~barns}$ or lower~\citep[this is consistent with the results of][who find hints of the onset of PISN near $70M_{\odot}$ from GWTC-3 using non-parametric mass-models]{MaganaHernandez:2025fkm}. The data prefers a smooth fall-off over a sharp gap for $m_2>40.0^{+9}_{-7}M_{\odot}$, a transition in the effective spin distributions near $m_1=43_{-5}^{+11}M_{\odot}$, and a post-transition sub-population which has $52^{+18}_{-23}\%$ events with mass-ratios in a broad range $(0.6,1)$. By comparing with simulations of dense stellar clusters~\citep{Rodriguez:2019huv, Kremer:2019iul}, we have shown that our post-transition mass-ratio distribution is difficult to explain exclusively with the theoretical predictions of (2G+1G) hierarchical mergers, and that contributions from (2G+2G) systems are not abundant enough to alleviate this discrepancy. Hence, for BBH formation above $40-50M_{\odot}$, alternative scenarios merit further consideration given the GWTC-4 detection sample.% The post-transition mass-ratio distribution is difficult to explain exclusively with the theoretical predictions for (2G+1G) hierarchical mergers in dense stellar clusters~\citep{Rodriguez:2019huv}. Furthermore, the low abundance of (2G+2G) systems relative to (2G+1G) in theoretical predictions indicates that alternative scenarios to hierarchical mergers dominating BBH formation above $\simeq 40M_{\odot}$ merit further consideration.

%By comparing with simulations of dense stellar environments, we have shown that our post-transition mass-ratio distribution is difficult to explain exclusively with the theoretical predictions of (2G+1G) hierarchical mergers, and that contributions from (2G+2G) systems are not abundant enough to alleviate this discrepancy.

In a recent study, \cite{Kiroglu:2025vqy} have shown that BH-star collisions and subsequent accretion-driven growth can lead to 1G+1G dynamical mergers in the $m_1\geq40M_{\odot}$ range. These systems are expected to be preferentially equal mass and comparable in abundance to hierarchical mergers. In Figure~\ref{fig:trans-acc}, we show that the presence of both hierarchical mergers and systems with accretion-grown components can be more consistent with our inferred distributions above the spin-transition mass than hierarchical mergers alone~\footnote{The data for these simulations from \cite{Kiroglu:2025vqy} were provided by Fulya K{\i}ro{\u{g}}lu~(fulyakiroglu2024@u.northwestern.edu).\label{fn:fulya}}. However, for this particular set of cluster simulations~\citep{Kiroglu:2025vqy}, there remains a marginal inconsistency with our inferred distributions for the high mass-subpopulation. This can indicate additional contributions from other formation channels.

In BBHs emerging from isolated stellar binaries, certain evolutionary phases can lead to a high-mass, symmetric, and highly spinning sub-population above $m_1\gtrsim40M_{\odot}$. \cite{Briel:2022cfl, vanSon:2020zbk} show that super-Eddington accretion during mass-transfer onto the first-born BH can lead to high masses, and distributions of mass-ratios, and effective-aligned spins consistent with our findings of the $m_1\geq \tilde{m}$ sub-population. On the other hand, chemically homogeneous evolution of tight binary systems might also lead to a high-mass, high effective-aligned-spin merging BBH sub-population~\citep{deMink:2016vkw}. Alternatively, BBH components born through the collapse of population-three stars can explain our inferred $\chi_{eff}$, and $q$ distributions above the spin-transition mass~\citep{Tanikawa:2024mpj}. 

Even though these studies do not make direct predictions on $\chi_p$ distributions, it is generally believed that spin orientations of BBHs formed in isolation are preferentially aligned to the orbit, leading to suppressed precession, unless there are strong natal kicks and inefficient tidal-realignment~\citep{Kalogera:1999tq, Bavera:2020inc, Gerosa:2018wbw, Steinle:2022rhj}.  Further investigation and detailed binary evolution models are necessary to rigorously explore these possibilities in the context of our inferred distributions.  %\textcolor{red}{Is this correct? Does this section need more? Vicky, suggest papers?}%https://arxiv.org/pdf/2412.13318 top heavy IMF?.

Alternatively, a third scenario is plausible. A subpopulation with comparable contributions from the isolated channel and hierarchical mergers can be consistent with our inferred distributions. Depending on the rate of dynamically assembled mergers and the fall-off in the mass spectrum of BBHs formed in isolation, contributions from the (2G+1G) hierarchical mergers can become significant even below the PISN cut-off. Substantial contributions from both these channels can give rise to a unique sub-population of systems above a certain mass range that exhibits similar trends to our inferred distributions, namely, broad spin distributions with substantial fractions of both aligned and misaligned/precessing systems, as well as a higher mass-symmetry than expected from (2G+1G) mergers alone. This is consistent with the population inference of \cite{Antonini:2025ilj}, whose non-parametric analyses indicate that measurement uncertainties can allow anywhere between $0-60\%$ of the systems above $m_1>\sim40M_{\odot}$ to be slowly spinning with preferentially aligned orientations. Observational confirmation of this hypothesis from future catalogs can lead to constraints on the relative abundance between BBHs formed in dynamical and isolated environments. Disentangling the contributions of multiple formation channels, whose relative abundance is expected to vary across cosmic time~\citep{Zevin:2020gbd}, will also necessitate flexible modeling of the redshift evolution of various features that characterize the high mass subpopulation~\citep[see, for example,][who report evidence for population level correlations between redshift and other binary properties in the past and current detection samples]{Biscoveanu:2022qac, Tenorio:2025nyt, Rinaldi:2025emt, Afroz:2025typ}.%of various population features identified in this analysis

%likely necessitate flexible modeling of the redshift evolution of the merger rate in addition to that of the distributions of intrinsic BBH parameters. The relative abundance of various formation mechanisms are exp

Nevertheless, due to broad measurement uncertainties, the inferred effective precessing spin distributions do not necessarily rule out the hypothesis that isolated BBH formation contributes dominantly (or substantially) above $40M_{\odot}$. Furthermore, the exact nature of the post-transition spin distributions might be susceptible to modeling assumptions. More data and non-parametric inference of the joint distribution of BBH masses and effective spins might be necessary to fully characterize this high-mass sub-population and rigorously establish its origin. Similarly, it is unclear from the current detection sample whether a PISN cut-off still exists around masses $\sim60M_{\odot}$ or higher. Updated GW catalogs at the end of LVK's ongoing fourth observing run~\citep{KAGRA:2013rdx} might elucidate these uncertainties and enable robust, model-independent constraints on PISN and stellar evolution theory. Novel insights into the origins of various high-mass BBH sub-populations are plausible. To avoid model-induced biases, the use of flexible parametrizations that do not explicitly model sharp features in the distribution will likely be crucial.
\vspace{-0.095cm}
%We note that due to broad measurement uncertainties, the inferred effective precessing spin distribution does not necessarily rule out these sub-channels of isolated BBH formation above $40M_{\odot}$. Furthermore, the exact nature of the post-transition spin distributions might be susceptible to modeling assumptions. More data and non-parametric inference of the joint distribution of BBH masses and effective spins might be necessary to fully characterize this high-mass sub-population and rigorously establish its origin. Similarly, it is unclear from the current detection sample whether a PISN cut-off still exists around masses $\sim60M_{\odot}$ or higher. Updated GW catalogs at the end of LVK's ongoing fourth observing run might elucidate these uncertainties and enable robust, model-independent constraints on PISN and stellar evolution theory. Novel insights into the origin of various BBH sub-populations are likely imminent.
\section{Acknowledgements}
We thank Fulya K{\i}ro{\u{g}}lu, Mike Zevin, Max Briel, Tassos Fragos, Ish Gupta, Ignacio Maga\~na Hernandez, Maya Fishbach, and Darsan Bellie for insightful discussions and suggestions. We are grateful to F.K. also for providing the data from their simulations. A.R. was supported by the National Science Foundation~(NSF) award PHY-2207945. V.K. was supported by the Gordon and Betty Moore Foundation (grant awards GBMF8477 and GBMF12341), through a Guggenheim Fellowship, and the D.I. Linzer Distinguished University Professorship fund. We are grateful for the computational resources provided by the LIGO laboratory and supported by National Science Foundation Grants PHY-0757058 and PHY-0823459. This material is based upon work supported by NSF’s LIGO Laboratory, which is a major facility fully funded by the National Science Foundation. This research has made use of data obtained from the gravitational Wave Open Science Center (gwosc.org), a service of LIGO Laboratory, the LIGO scientific Collaboration, the Virgo Collaboration, and KAGRA. We gratefully acknowledge the support of the NSF-Simons AI-Institute for the Sky (SkAI) via grants NSF AST-2421845 and Simons Foundation MPS-AI-00010513.

\appendix

\section{Distribution functions}
\label{secc:appendix-dist}
The exact functional forms of the various components of the our mass-ratio model are given by:
\begin{equation}
    \mathcal{TN}(q, m_1, \mu_{m_2} \sigma_{m_2}, m_{2,b2}) = \begin{cases}
        \frac{m_1e^{-\frac{1}{2}\left(\frac{m_1q-\mu_{m_2}}{\sigma_{m_2}}\right)^2}}{\sigma_{m_2}}\times \left(\Phi\left(\frac{m_1-\mu_{m_2}}{\sigma_{m_2}}\right)-\Phi\left(\frac{m_{2,b2}-\mu_{m_2}}{\sigma_{m_2}}\right)\right)^{-1}, & m_{2,b2}\leq qm_1 \leq m_1,\\
        0, & o.w.
    \end{cases}
\end{equation}
is a truncated Gaussian with $\Phi$ being the CDF of the standard normal distribution, and
\begin{equation}
    \mathcal{BPL}(q, m_1, m_{2,b1}, \beta_1, \beta_2, m_{2,min}) = \begin{cases}\frac{q^{\beta_1}(1+\beta_1)}{1-(\frac{m_{2,min}}{m_1})^{\beta_1+1}}, & m_{2,min}<q  m_1 < m_1 < m_{2,b1}\\
       \frac{q^{\beta_1}\left(\frac{m_{2,b}}{m_1}\right)^{(\beta_2 - \beta_1)}(1+\beta_2)}{1-(\frac{m_{2,b1}}{m_1})^{\beta_2+1}} , & m_{2,min}<q  m_1 <  m_{2,b1} < m_1\\
       \frac{q^{\beta_2}(1+\beta_2)}{1-(\frac{m_{2,b1}}{m_1})^{\beta_2+1}}, & m_{2,b1}<q  m_1 < m_1 \\
       0, & o.w.\end{cases}
\end{equation}
is a \textit{continuous} broken power-law distribution. Here, $m_{2,b1}$ is the break location in the powerlaw, parametrized in terms of the secondary mass, $\beta_1$, and $\beta_2$ the powerlaw indices below and above the break, respectively, $m_{2,min}$ is the minimum of the secondary mass. The mean $(\mu_{m_2})$, standard deviation~$(\sigma_{m_2})$, and lower bound of the truncated Gaussian are also parametrized in units of secondary mass.

For the effective spin distributions, the functional forms of the truncated Gaussian components are given by:
\begin{equation}
    \mathcal{TN}_{\chi}(\chi,  \mu_{\chi} \sigma_{\chi}, \chi_{min}, \chi_{max}) = \begin{cases}
        \frac{e^{-\frac{1}{2}\left(\frac{\chi-\mu_{\chi}}{\sigma_{\chi}}\right)^2}}{\sigma_{\chi}}\times \left(\Phi\left(\frac{\chi_{max}-\mu_{\chi}}{\sigma_{\chi}}\right)-\Phi\left(\frac{\chi_{min}-\mu_{\chi}}{\sigma_{\chi}}\right)\right)^{-1}, & \chi_{min}\leq \chi \leq \chi_{max},\\
        0, & o.w.
    \end{cases}
\end{equation}
where $\mu_{\chi}, \sigma_{\chi}, \chi_{\min}, \chi_{max}$ are the mean, standard deviation, minimum and maximum of the truncated Gaussian respectively.

\section{Hyper-priors}
\label{sec:appendix-priors}
In this section, we list the priors on our model hyperparameters that were used in the inference presented in the main text. For the mass-ratio distributions, the hyperpriors on the new parameters introduced in our models are shown in table~\ref{table:priors-mass}. Hyper-priors on the spin-transition models are shown in table~\ref{table:priors-spin}. For the components in our population model that are identical to the default distributions of \cite{LIGOScientific:2025pvj}, we use the same hyperpriors as theirs.
\begin{table}[h]
\centering
\begin{tabular}{cccc}
\hline
\hline
Hyperarameter & Flexible & Gap-enforced & Broad\\
\hline
$\beta_2$ & $U(-20,\mathbf{3})$ & $ U(-\mathbf{140}, 70)$ & $\mathbf{U(-200, 0)}$\\
$m_{2,b1}$ & $U(20M_{\odot},60M_{\odot})$ & $U(20M_{\odot},60M_{\odot})$ & $\mathbf{U(20M_{\odot},60M_{\odot})}$\\
$m_{2,b2}$ & $U(m_{2,b1},120M_{\odot})$ & $U(100M_{\odot},200M_{\odot})$ & $\mathbf{U(m_{2,b1},120M_{\odot})}$ \\
$\mu_{m_{2}}$ & $U(m_{2,b2},300M_{\odot})$ & $U(m_{2,b2},300M_{\odot})$ & $\mathbf{U(m_{2,b2},300M_{\odot})}$\\
$\sigma_{m_{2}}$ & $U(0,200M_{\odot})$ & $U(0,200M_{\odot})$ & $\mathbf{U(0,200M_{\odot})}$\\
$\lambda_q$ & $U(0,1)$ & $ U(0, 1)$& $\mathbf{U(0, 1)}$\\
\hline
\end{tabular}
\caption{Hyperpriors on the additional parameters introduced by our mass-ratio distribution.}
\label{table:priors-mass}
\end{table}
\begin{table}[h]
\centering
\begin{tabular}{cc}
\hline
\hline
Hyperarameter & Prior \\
\hline
$\mu_{\chi_{eff},1}$ & $U(-1,1)$ \\
$\mu_{\chi_{eff},2}$ & $U(-1,1)$ \\
$\sigma_{\chi_{eff},1}$ & $U(0.05,1)$ \\
$\sigma_{\chi_{eff},2}$ & $U(0.05,1)$ \\
$\mu_{\chi_{p},1}$ & $U(0.05,1)$ \\
$\mu_{\chi_{p},2}$ & $U(0.05,1)$ \\
$\sigma_{\chi_{p},1}$ & $U(0.05,1)$ \\
$\sigma_{\chi_{p},2}$ & $U(0.05,1)$ \\
$\tilde{m}$ & $U(20M_{\odot},60M_{\odot})$\\
\hline
\end{tabular}
\caption{Hyperpriors on the parameters of our effective spin distributions.}
\label{table:priors-spin}
\end{table}

\section{Measurability of Percentiles}
\label{sec:appendix-m299}
\begin{figure*}
    \centering
    \includegraphics[width=0.8\linewidth]{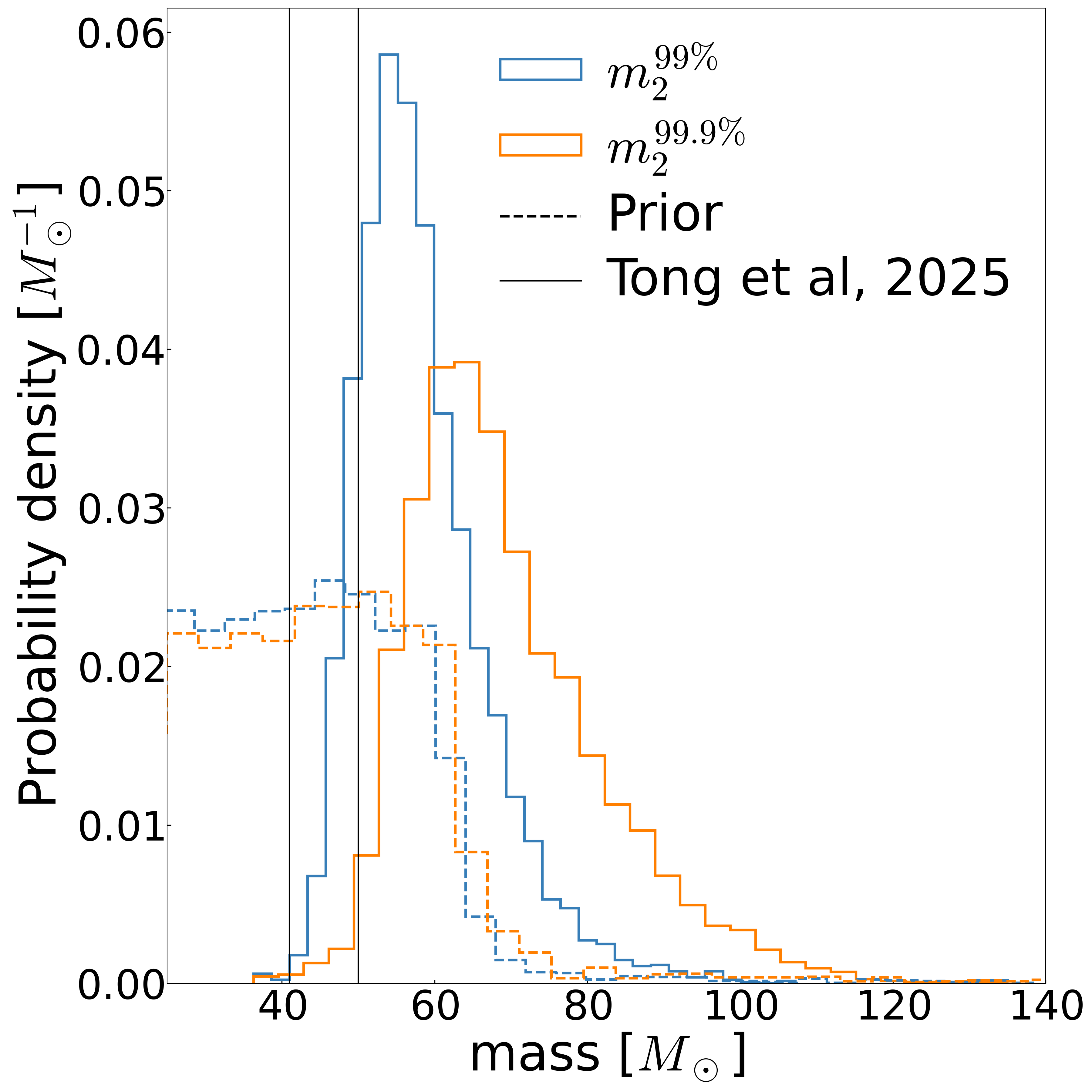}
    \caption{The posterior distribution of the $99$th, and $99.9$th percentiles of the astrophysical distribution in $m_2\in (m_{2,b1}, m_{2,b2})$, compared with the corresponding priors.}
    \label{fig:m299wprior}
\end{figure*}
Here, we explore the prior-dependence of our measurement of $m_2$ percentiles, which is relevant to our constraints on the lower bounds of possible gap onset that is still allowed within measurement uncertainties. In Figure~\ref{fig:m299wprior}, we compare the posterior distribution of $m_{2}^{99}$ and $m_2^{99.9}$ compared with their prior predicitive distributions. It can be seen that the data clearly informs these measurements much more than the prior, despite the restriction $m_2\in(m_{2,b1}, m_{2,b2})$.

\section{Dependence on the Primary mass distribution model}
\label{sec:appendix-pm1}
\begin{figure*}[htt]
    \centering
    \includegraphics[width=0.98\linewidth]{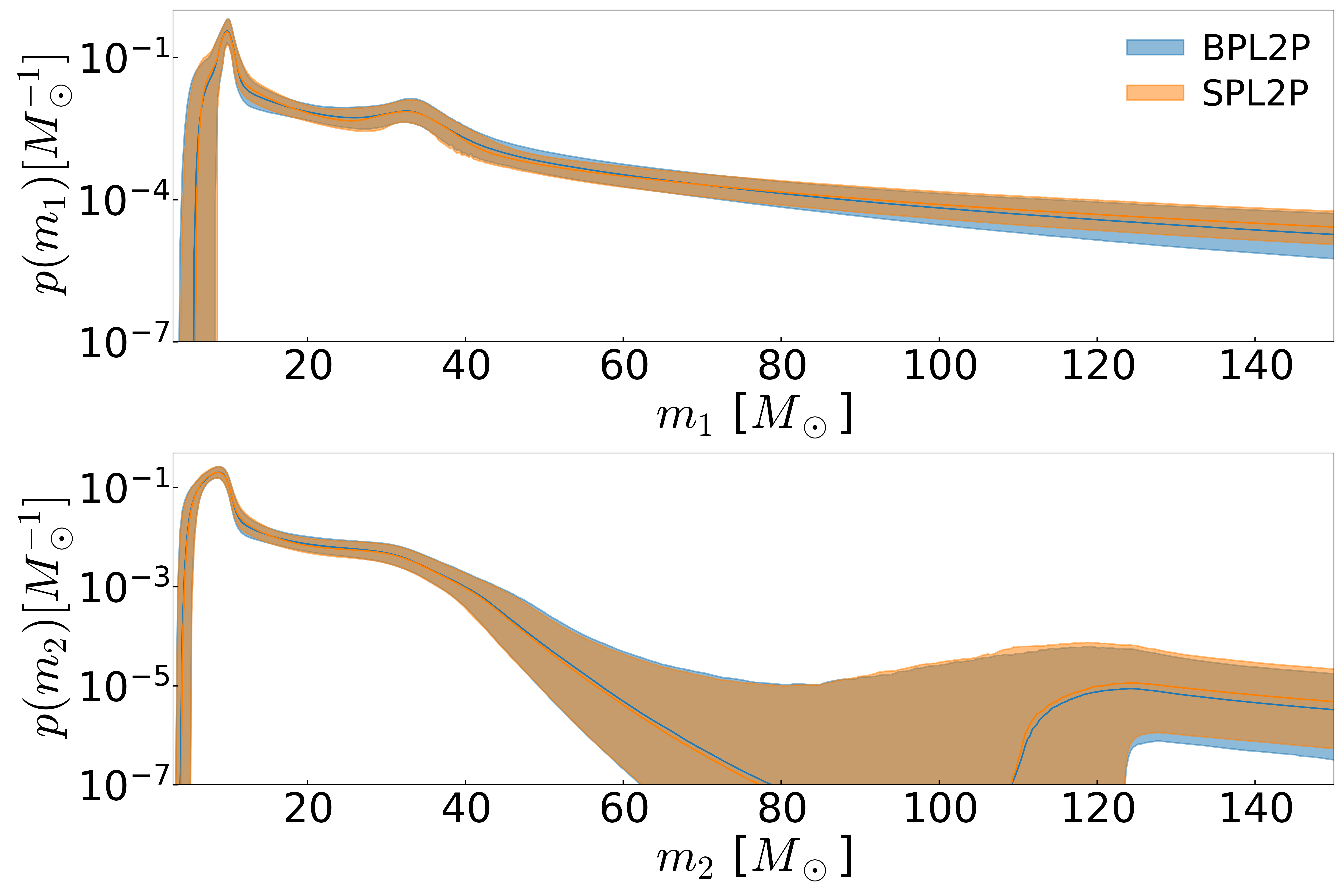}
    \caption{Inferred mass distribution of primary (top) and secondary (bottom) BH components.}
    \label{fig:primary-diff}
\end{figure*}
Here we show our inferred mass distributions for the two different $p(m_1)$ models, namely the \texttt{BPL2P} and the \texttt{SPL2P}. For the mass-ratio distribution, we use the model of Eq.~\eqref{eq:q-dist} as in the main text, with flexible priors. In Figure~\ref{fig:primary-diff}, we show that our inferred distributions are consistent between the two different choices for $p(m_1)$. In conjunction with our flexible mass-ratio model, the \texttt{SPL2P} case is marginally preferred by the data over the \texttt{BPL2P} with a Bayes factor of 1.7.

\section{Comparison with the latest Cluster catalog}
\label{sec:appendix-cmc}
\begin{figure*}[htt]
    \centering
    \includegraphics[width=0.31\linewidth]{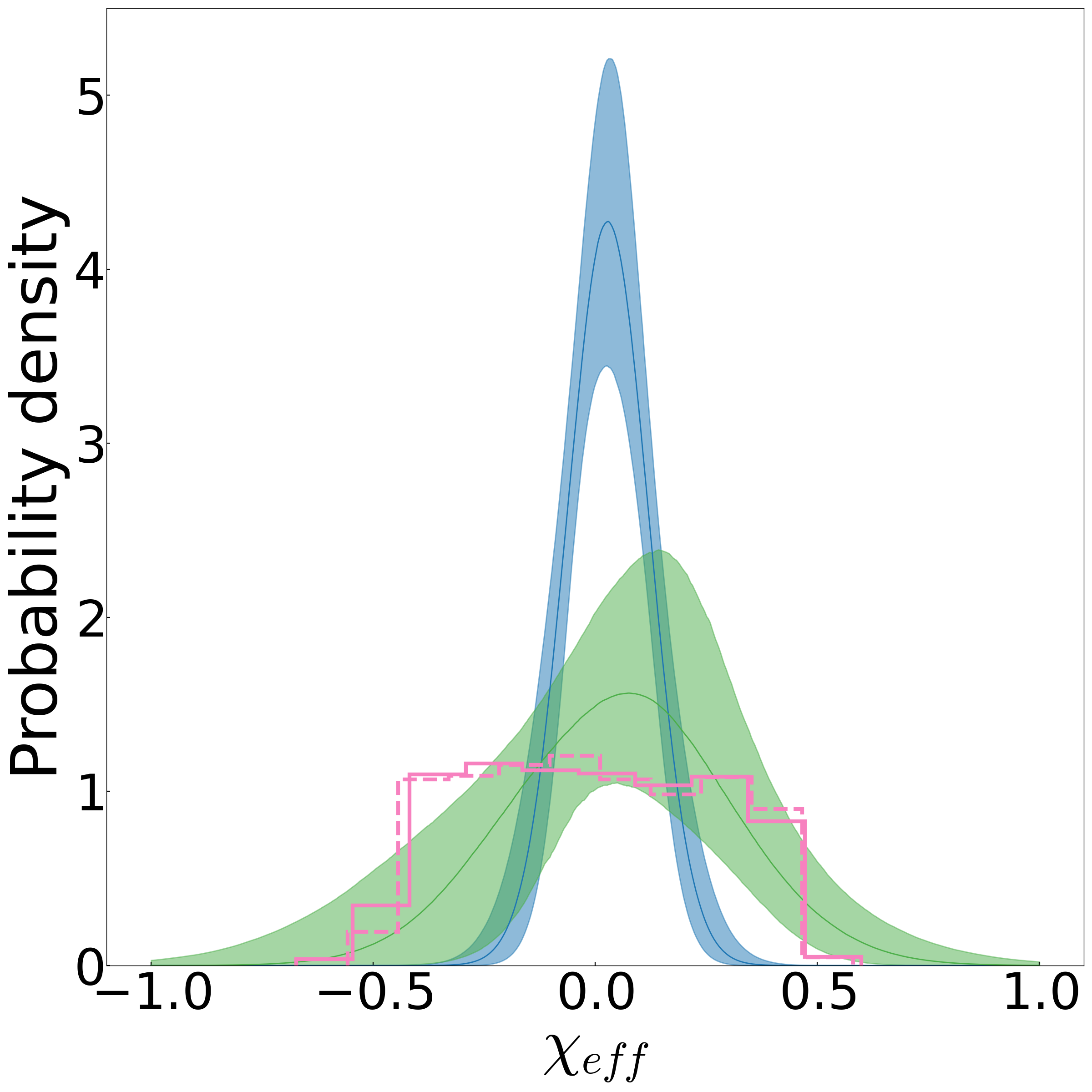}
    \includegraphics[width=0.31\linewidth]{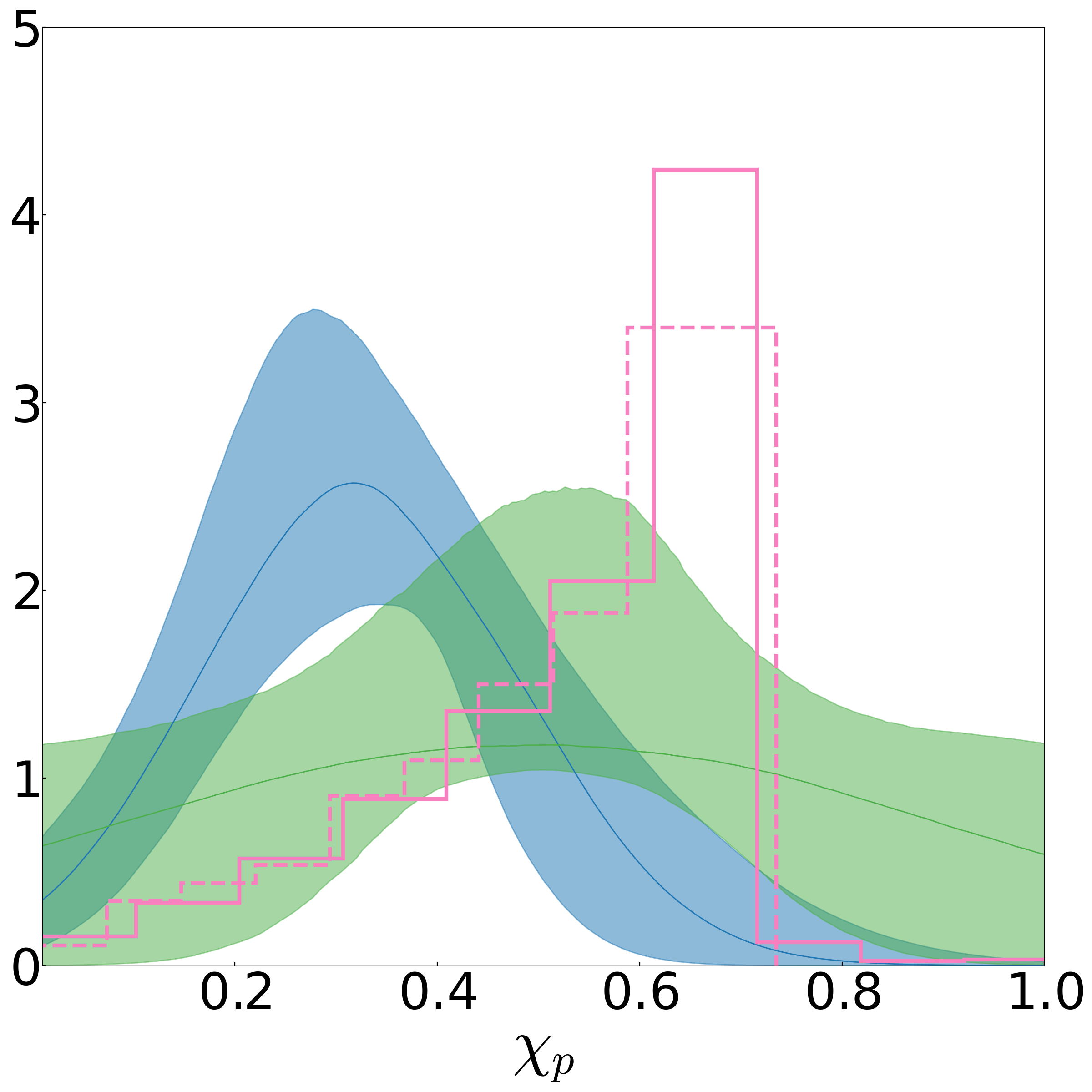}
    \includegraphics[width=0.31\linewidth]{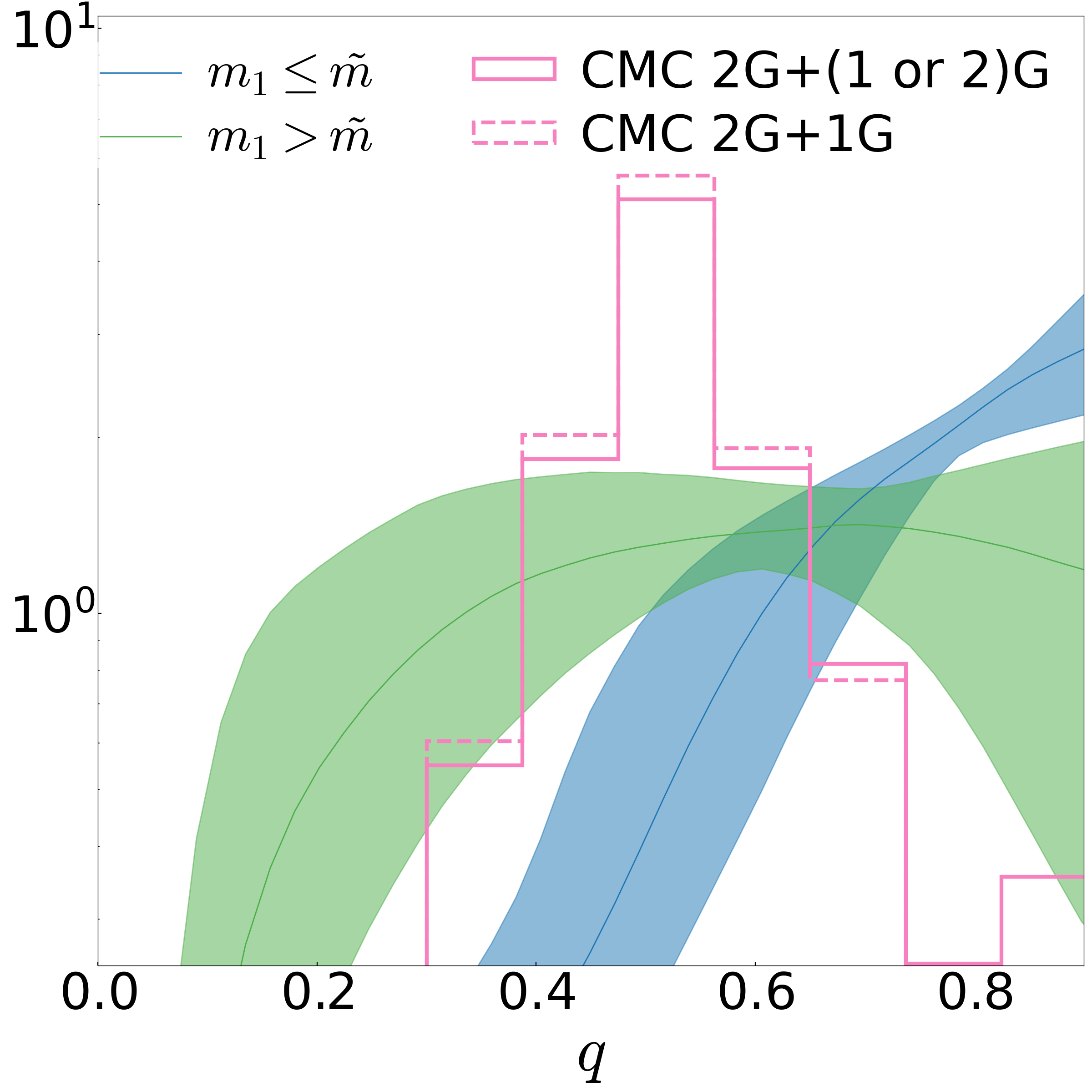}
    \caption{Comparing the post-spin-transition mass-ratio distribution with the latest public CMC catalog.}
    \label{fig:cmc-cat}
\end{figure*}

In this section, we compare our inferred mass-ratio distributions above the spin-transition mass with the latest catalog of CMC simulations~\citep{Kremer:2019iul}. We give equal weights to all catalogs, similar to~\cite[e.g.,][]{Borchers:2025sid}, and obtain the same inconsistency with our post-transition mass-ratio distribution as with the simulations of \cite{Rodriguez:2019huv}, which we show in Figure~\ref {fig:cmc-cat}. We have verified that scaling the contributions of each cluster by the inverse of the squared cluster mass makes no noticeable difference in the predicted distributions. Note, however, that BBHs in the latest public \textsc{CMC} catalog are not weighted by the fraction of clusters per metallicity at different redshifts, and their merger times are not convolved with the distribution of cluster formation time across different metalicities. In other words, these samples are not representative of the underlying distribution of dynamically assembled BBHs merging across cosmic time,  unlike the dataset released by \cite{Rodriguez:2019huv} as part of \cite{michael_zevin_2020_4277620} wherein the mentioned weighting schemes were implemented. For this reason, we have chosen to compare with the simulations of \cite{Rodriguez:2019huv} in the main text and discuss the latest catalog in this appendix. Nevertheless, our conclusions remain the same. 

\section{Posterior predictive checks and re-weighted event posteriors}
\label{sec:appendix-ppd}
\begin{figure*}[htt]
    \centering
    \includegraphics[width=0.98\linewidth]{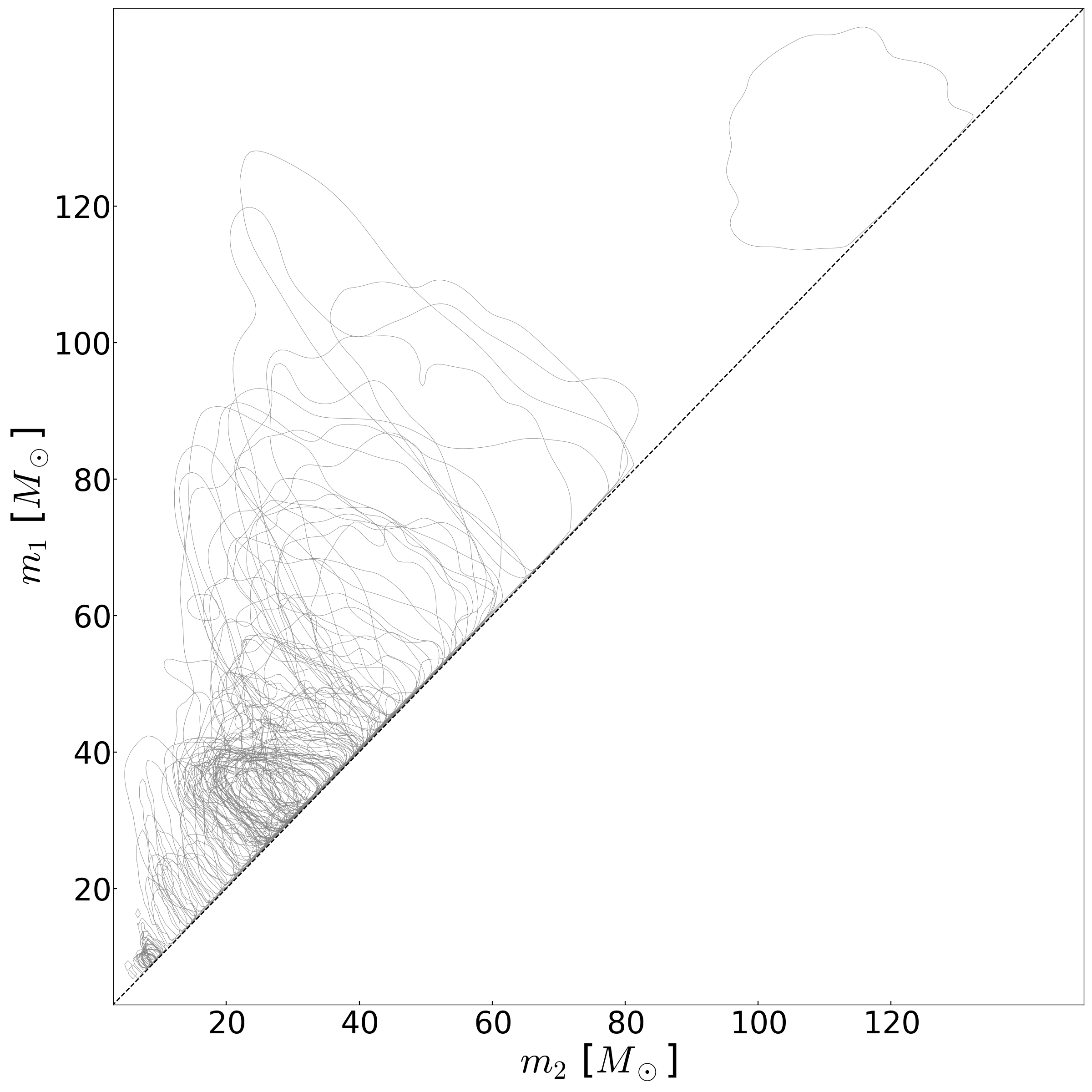}
    \caption{Component masses of individual events re-weighted by the population informed prior.}
    \label{fig:reweight-m1m2}
\end{figure*}
\begin{figure*}[htt]
    \centering
    \includegraphics[width=0.98\linewidth]{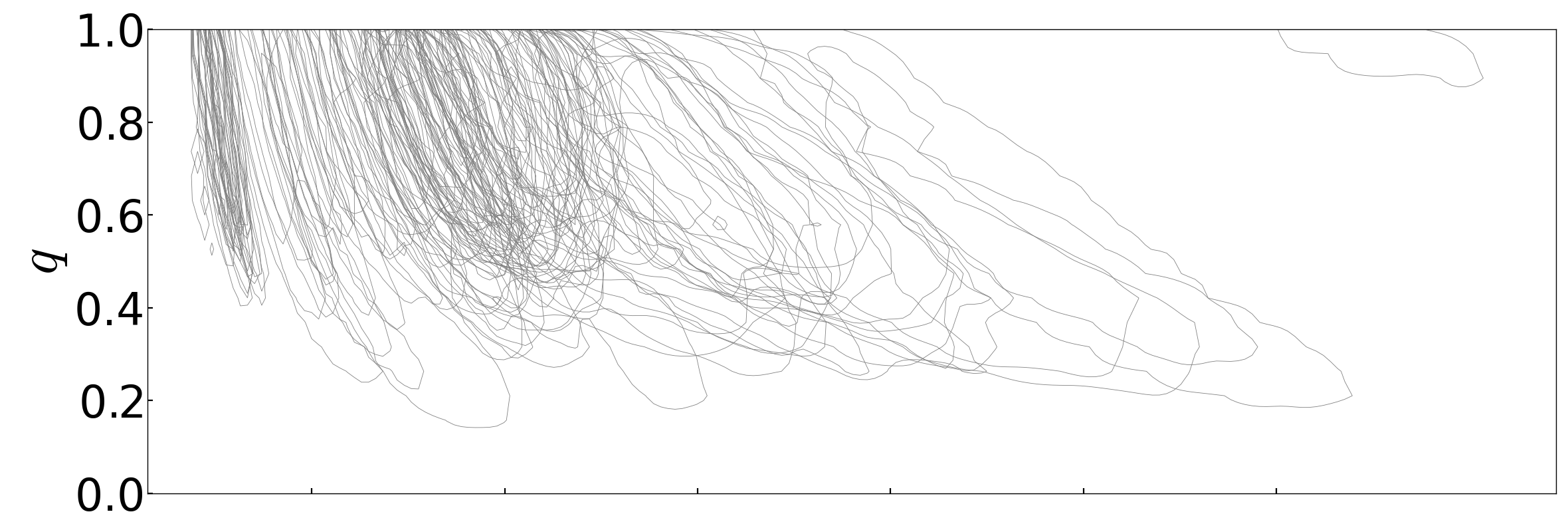}
    \includegraphics[width=0.98\linewidth]{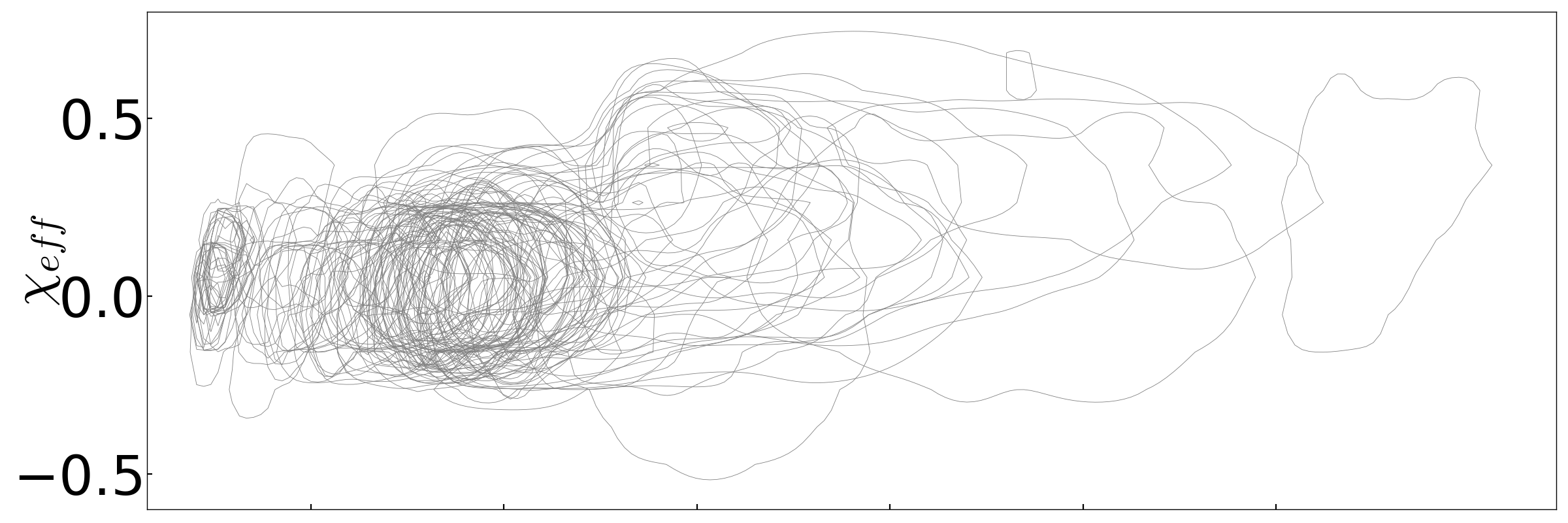}
    \includegraphics[width=0.98\linewidth]{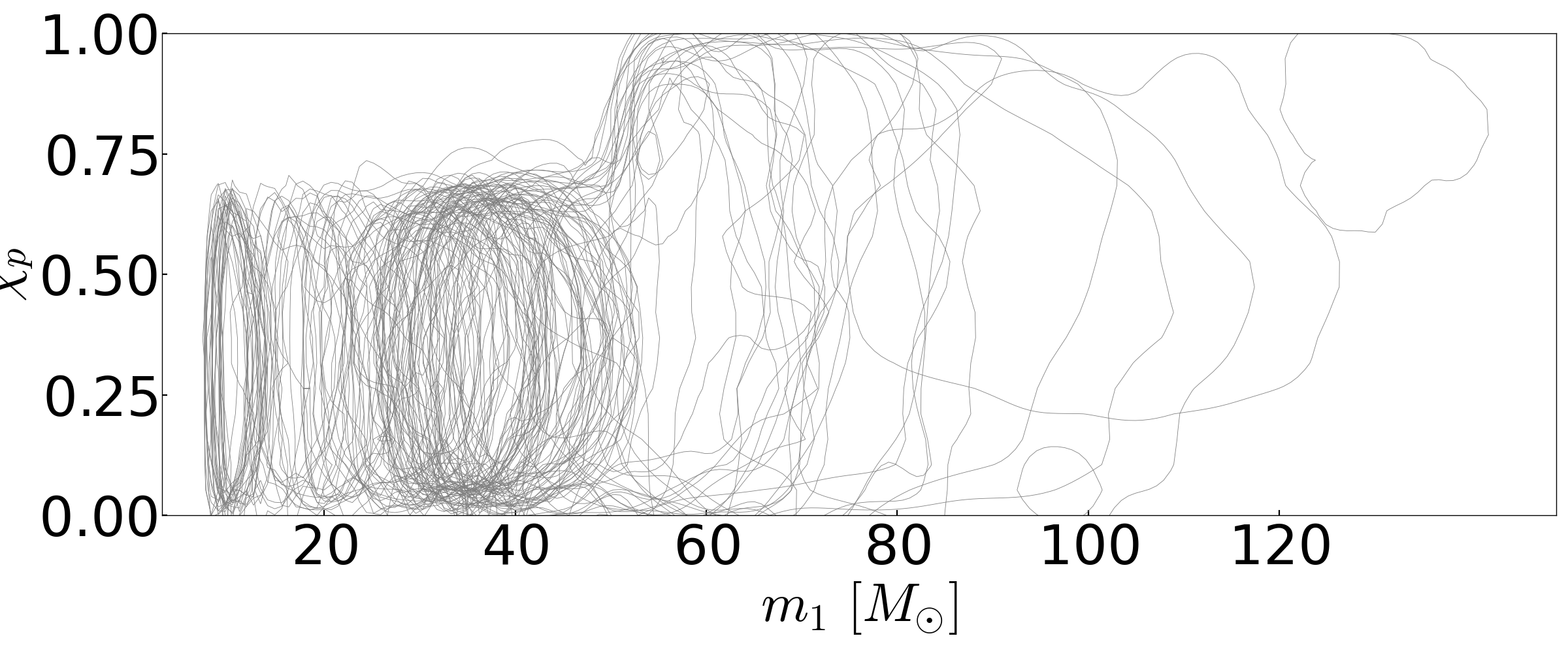}
    \caption{Primary mass, mass-ratio and effective spins of individual events re-weighted by the population informed prior.}
    \label{fig:reweight-others}
\end{figure*}

Here, we perform posterior predictive checks and test the goodness of fit for our best-performing model~\citep{Fishbach:2019ckx, Callister:2022qwb, Miller:2024sui}. We first re-weight the posterior distributions of each event to a population-informed prior obtained from our inferred distributions. Figure~\ref{fig:reweight-m1m2} shows the component masses of reweighted events in our detection sample and exhibits no evidence for a sharp gap in $m_2$ near $40-50M_{\odot}$. Figure~\ref {fig:reweight-others} shows the joint posterior distribution of primary mass, mass-ratio, and effective spins for each re-weighted event. It can be seen that there is a significant fraction of events with symmetric mass ratios above the mass range where the spin distributions start to broaden. In other words, there is no evidence of an observed high-mass sub-population in GWTC-4 that comprises exclusively 2G+1G hierarchical systems.
% as well as the set of detectable injections with a population-informed prior obtained from our inferred model. The reweighted events yield the observed distribution, whereas the reweighted injections yield the predicted one. The two are compared in Figure~\ref{fig:ppd}. The traces are found to be clustered along the diagonal, which indicates no significant tension between the fitted model and the observed population. For further details regarding how these traces are computed from the reweighted samples see appendix E of .

\begin{figure*}[htt]
    \centering
    \includegraphics[width=0.48\linewidth]{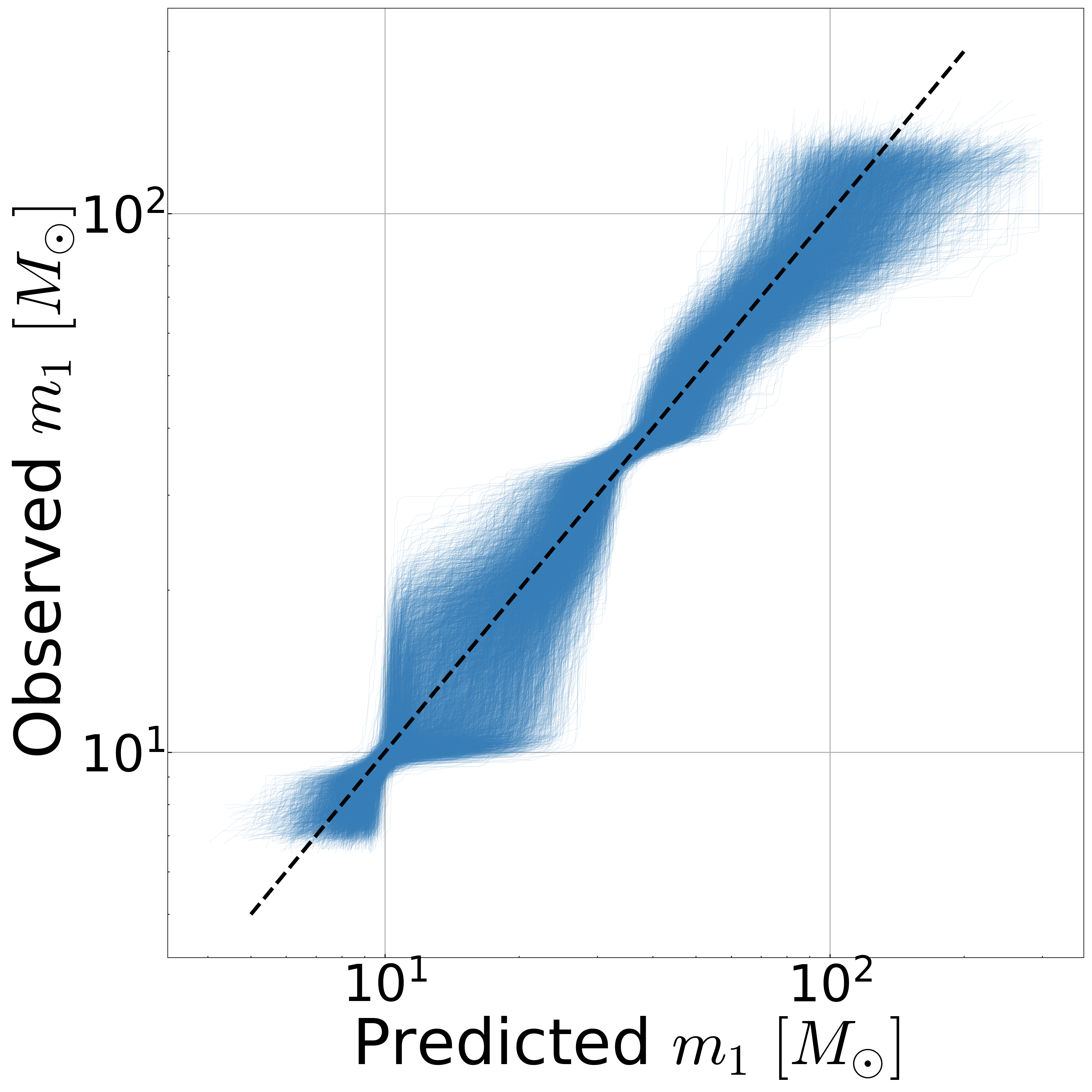}
    \includegraphics[width=0.48\linewidth]{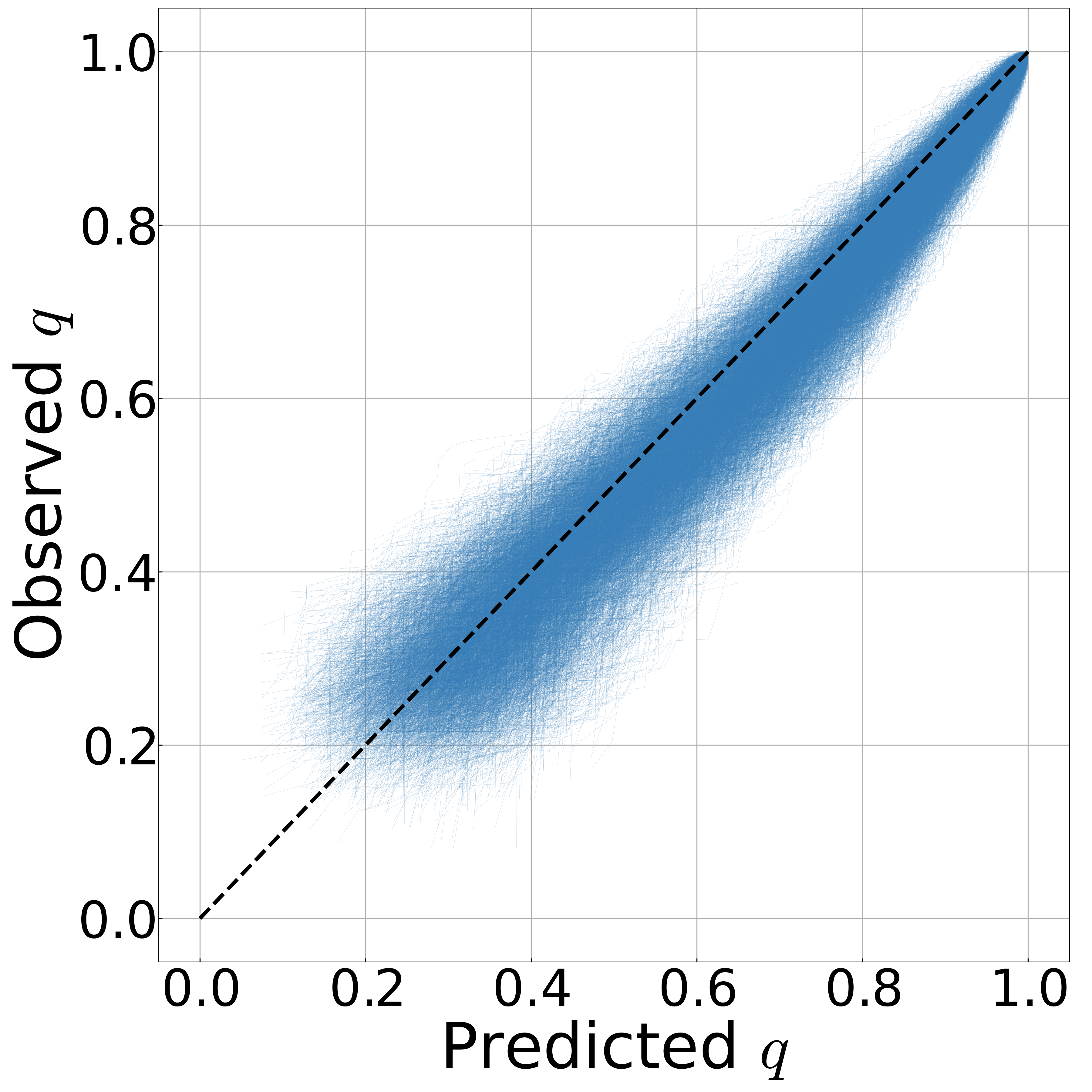}
    \includegraphics[width=0.48\linewidth]{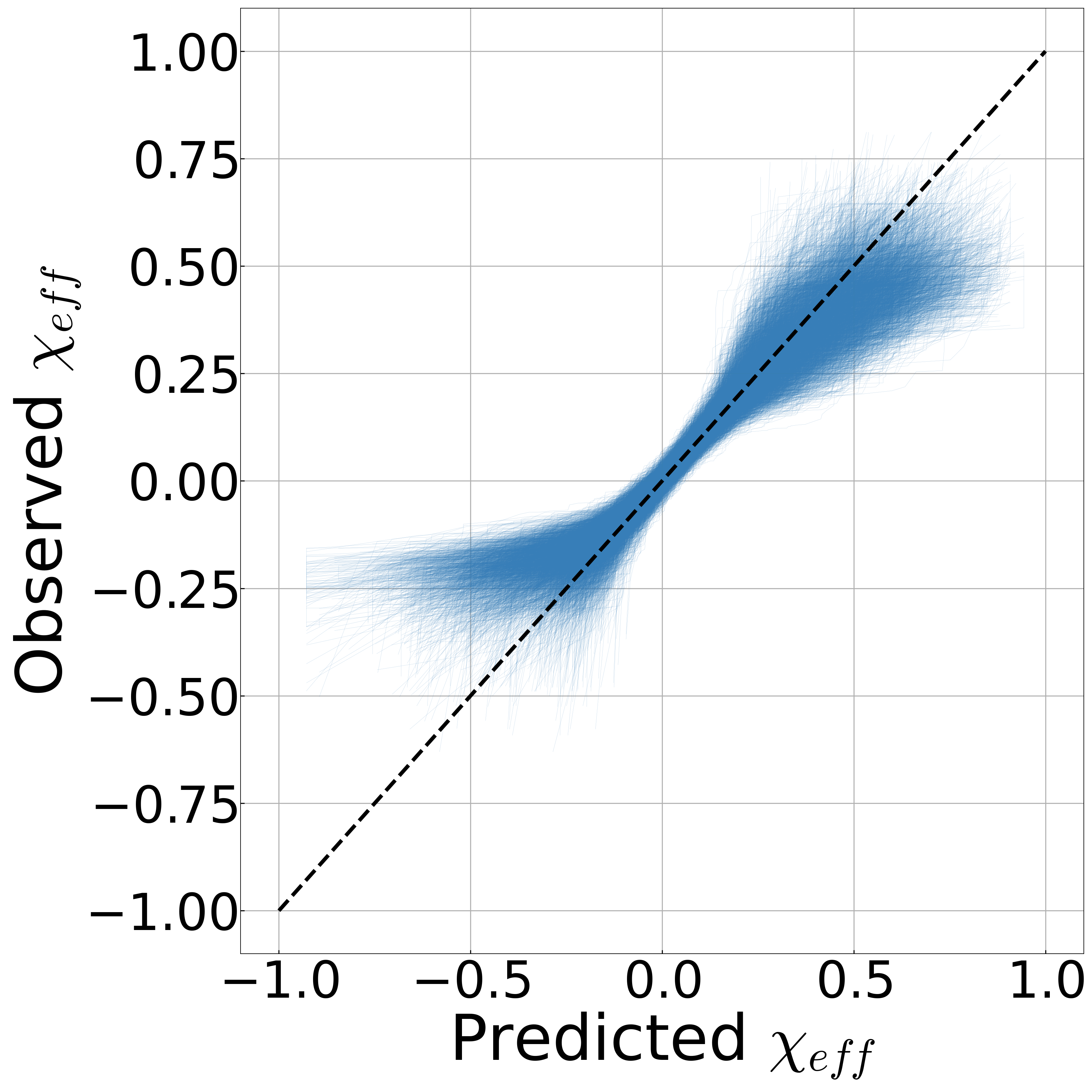}
    \includegraphics[width=0.48\linewidth]{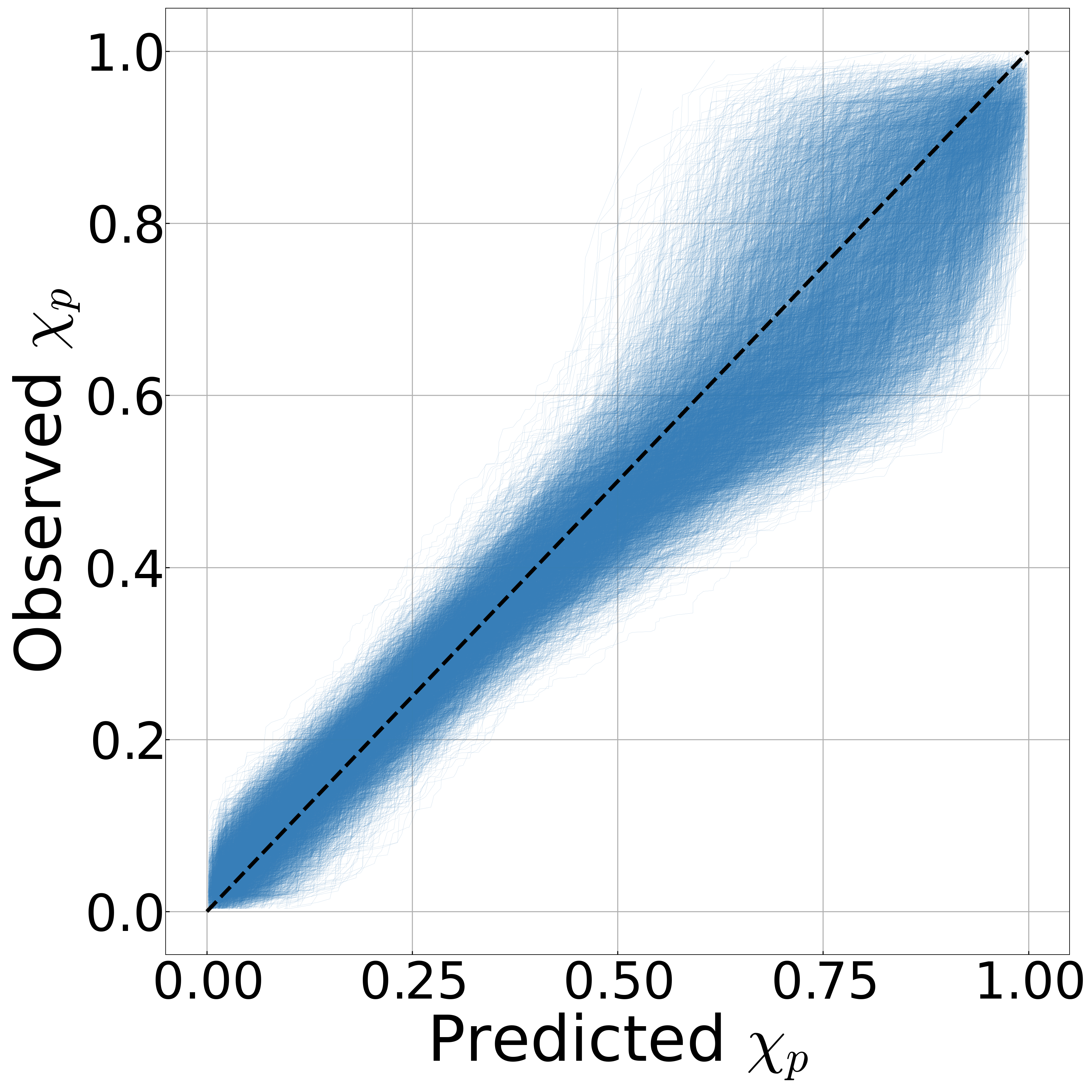}
    \caption{Posterior predictive checks for the inferred component-mass and effective spin distributions.}
    \label{fig:ppd}
\end{figure*}

Next, we perform the posterior predictive test for the goodness of fit of our inferred model. We reweight the set of detectable injections to the population informed prior and compute the predicted distribution of detections. We compare the predicted distribution with the observed one obtained from re-weighting individual detections to the same population-informed prior. Figure~\ref{fig:ppd} shows the results of this comparison and demonstrates that there is no systematic miss-modeling in the inference of our model from the observed events. See appendix E of~\cite{Callister:2023tgi} for details of how the trace plots are computed from the data and the hyper-posterior samples of the constrained population model.

% We further show the parameters of our re-weighted events in Figure~ which demonstrates that there is no clear evidence of a sharp gap near $40-50M_{\odot}$ in the GWTC-4 detection sample.

%\section{Hyper-posteriors}
%\label{sec:appendix-posteriors}
\bibliography{sample701}{}
\bibliographystyle{aasjournalv7}



\end{document}